\documentclass[aps,pra,showpacs,twocolumn, epsfig]{revtex4}
\usepackage[dvips]{graphics}
\usepackage{graphicx}
\usepackage{amsfonts}
\usepackage{amssymb}
\usepackage{amsmath}
\begin{document}
\title{Gap-Townes solitons and localized excitations in low dimensional
Bose-Einstein condensates in optical lattices}
\author{Fatkhulla Kh.Abdullaev}
\affiliation{Dipartimento di Fisica \textquotedblleft E.R.
Caianiello", Universit\'{a} di Salerno, I-84081 Baronissi (SA),
Italy, and Physical-Technical Institute of the Academy of
Sciences, 700084, Tashkent-84, G.Mavlyanov str.,2-b, Uzbekistan}
\author{Mario Salerno}
\affiliation{Dipartimento di Fisica \textquotedblleft E.R.
Caianiello", \\ Consorzio Nazionale Interuniversitario per le
Scienze Fisiche della Materia (CNISM), \\Istituto Nazionale di
Fisica Nucleare (INFN) Sezione di Napoli-Gruppo Collegato di
Salerno, Universit\'{a} di Salerno, I-84081 Baronissi (SA), Italy}
\thanks{E-mail: fatkh@physic.uzsci.net}
\thanks{E-mail: salerno@sa.infn.it}
\begin{abstract}
We discuss localized ground states of Bose-Einstein condensates in
optical lattices with attractive and repulsive three-body
interactions in the framework of a quintic nonlinear Schr\"odinger
equation which extends the Gross-Pitaevskii equation  to the one
dimensional case. We use both a variational method and a
self-consistent approach to show the existence of unstable
localized excitations which are similar to Townes solitons of the
cubic nonlinear Schr\"odinger equation in two dimensions. These
solutions are shown to be located in the forbidden zones of the
band structure, very close to the band edges, separating decaying
states from stable localized ones (gap-solitons) fully
characterizing their delocalizing transition. In this context
usual gap solitons appear as a mechanism for arresting collapse in
low dimensional BEC in optical lattices with attractive real
three-body interaction. The influence of the imaginary part of the
three-body interaction, leading to dissipative effects on gap
solitons and the effect of atoms feeding from the thermal cloud
are also discussed. These results may be of interest for both BEC
in atomic chip and Tonks-Girardeau gas in optical lattices.
\end{abstract}
\pacs{03.75.Lm;03.75.-b;05.30.Jp}
\date{\today }
\maketitle
\section{Introduction}
Bose-Einstein condensates (BEC) in periodic potentials are
presently receiving a great deal  of theoretical and experimental
interest due to the possibility to explore a whole class of
phenomena ranging from Bloch oscillations \cite{And,Cris},
Landau-Zener tunneling \cite{Wu} and solitons \cite{Eier} to
quantum phase transitions of the Mott insulator type \cite{Gren}.
In the mean field approximation these systems are described by the
Gross-Pitaevskii equation (GPE) which is a cubic nonlinear
Schr\"odinger (NLS) equation with periodic potential in which the
cubic nonlinearity models the two-body interatomic interactions
appropriate for dilute gases. The presence of the optical lattice
(OL) allows interesting localization phenomena such as the
formation of gap-solitons, i.e. localized states with energies
comprised in the gaps of the band structure of the underlying
linear periodic problem, both for attractive and repulsive
interactions \cite{Baiz,Ostr}. This is quite remarkable because it
is known that for two (2D) and three dimensional (3D) NLS, in
absence of periodic potential, soliton solutions do not exist and
for attractive interactions the phenomenon of collapse in finite
time appears. In this last case only one localized solution is
possible, the so called Townes soliton, existing for a single
value of the norm (number of atoms). This solution however  is
unstable against norm variations i.e. it collapses in a finite
time for norms above the critical value and decays into the
uniform background for norms below the critical value. The
presence of the OL allows to expand the range of existence of
localized solutions from a single value to a whole interval of
norms below which the delocalizing transition occurs and above
which collapse takes place.

A description based on the GPE with cubic nonlinearity, however,
is adequate only at low densities. For higher densities the
three-body interactions start to play a role and a description
based on two-body interactions is no longer sufficient.
\cite{Akhmediev,Zhang,AGTF,GFTA}. To this regard we recall that
condensate densities limited by three-body inelastic collisions
\cite{ketterle} have indeed been achieved. In these regime a more
accurate treatment of the mean-field energetics of a dense
condensate will need to account for both two- and three-body
elastic collisions.

The contribution of the three-body interactions can be enhanced by
detuning to zero the cubic two body term by means of Feschbach
resonances \cite{ketterle}, this leading to a periodic NLS
equation with quintic nonlinearity. For appropriate values of
density and scattering lengths, however, the three-body collisions
could largely dominate the two-particle contributions even in a
very dilute regime. This occurs when the so called Efimov effect
\cite{efimov} becomes possible i.e. when the two-body scattering
length becomes much larger than the effective two body interaction
radius (this usually occurring near a two-body resonance). In this
case a very large number of three-body bound states can be formed
in the system and the contribution of the three-body elastic
collisions (cubic in the density) to the energy may largely
overcome the one arising from the two-body terms (quadratic in the
density). In this situation the condensate appear extremely dilute
with respect to the two-body collisions but somewhat dense with
respect to the three-body collisions \cite{bulgac}.

Theoretical estimates of the three-body coefficient were given in
\cite{Bed,Braat,Jack,Koh}. This term can be modeled by a quintic
nonlinearity in the GPE which for rubidium atoms  is expected to
be attractive \cite{Zhang}.  The different types of instabilities
induced by the quintic term may lead to interesting dynamical
phenomena in 1D BEC similar to Bose-Novae effect observed in
multidimensional case. On the other hand,  the presence of these
instabilities restricts the possibility of BEC manipulations in
atomic waveguides \cite{Zhang,AGTF}. The optical lattice, however,
can suppress or delay some of these instabilities opening the
possibility for new types of localized excitations.

The 1D quintic NLS is also used to describe a Bose gas with
hard-core interactions in the Tonks-Girardeau regime. Recent works
have shown that this approach describes well the ground state
properties, the collective oscillations in a parabolic trap
\cite{Minguzzi} and the dynamics of shock waves and dark solitons
in the gas \cite{Damski,Kevr}. On the other hand, interference
phenomena generated with small number of atoms ($N = 10$) were
shown to be not well accounted by the quintic NLS (the
interference patterns obtained from this equation are much more
pronounced than the experimental ones) \cite{Gir00}. For larger
number of atoms and for weak density modulations, however, the
description of the nonlinear excitations in the Tonks-Girardeau
regime by means of a quintic NLS equation is expected to be valid.

The aim of the  present paper is to study localized states of the
quintic nonlinear Schr\"odinger  equation with periodic potential
for both attractive and repulsive interactions. To this regard we
use a variational method, a self-consistent approach and direct
numerical integrations, to show that the presence of the optical
lattice allows to stabilize solitons of the attractive quintic 1D
NLS against collapse or decay. In the case of repulsive
interactions the optical lattice is found to be crucial for the
existence of stable bright matter waves. The existence of
localized excitations in this system is shown to be associated to
the existence of unstable localized solutions similar to the
Townes soliton of the cubic 2D NLS equation (we refer to them as
gap-Townes solitons). These unstable solutions are found to be
located in the forbidden zones of the band structure, very close
to band edges and separate decaying states from stable
gap-solitons. The existence curve of these solutions characterizes
the critical threshold for the occurrence of the delocalizing
transition and the existence of gap solitons appears to be a
mechanism for arresting collapse in 1D BEC in OL with three-body
interactions. We also investigate dissipative effects on gap
solitons induced by the imaginary part of the three-body
interactions. We show that when the imaginary part of this
interaction is small compared to the real part, localized states
exist for very long times and their behavior is qualitatively
similar to the one of the undamped case. For moderate and large
damping, however, all localized states in the band gap decay into
the Bloch state at the band edge. We find remarkable that even in
this case a reminiscence of the existence of gap-Townes solitons
survives in focusing-defocusing cycles observed when the number of
atoms is close to the critical value for their existence. We also
find that these localized states can be stabilized into
breather-like excitations  by a linear amplification term modeling
the feeding of atoms in the condensate from the thermal cloud in
the presence of nonlinear damping.

We remark that in absence of periodic potential the quintic 1D NLS
equation has a behavior similar to the 2D NLS equation with cubic
nonlinearity. The interplay between dimensionality and
nonlinearity has indeed been used in the past to investigate
collapse in lower dimensional NLS, the critical condition being
$D(n-1)-4=0$ where $n$ the order of the nonlinearity in the
equation and $D$ is its dimensionality \cite{Berge}. From this
point of view the quintic NLS can be viewed as a 1D model for the
2D GP with cubic mean field nonlinearity and we expect that the
results discussed in this paper  will apply also to this case.

The paper is organized as follows. In Section II we introduce the
1D model and discuss the range of applicability to different
physical situations. In Section III we study the quintic 1D NLS
equation with attractive interactions both in absence and in
presence of a periodic potential. The effectiveness of the optical
lattice to stabilize localized states of the GPE in presence of
two-body and three-body interactions is investigated by means of a
variational analysis and by direct numerical simulations. We use a
self-consistent method to investigate gap solitons and band
structures. The existence of gap-Townes solitons is shown in
Section IV and the role of usual gap solitons in arresting
collapse is discussed. In Section V we investigate the case of
repulsive interactions  by means of variational analysis,
self-consistent method and numerical simulations. The existence of
unstable Townes soliton with energies above the band edges and
their role in the delocalizing transition is also investigated. In
Section VI the dissipative effects introduced by an imaginary part
of the three-body interaction on gap solitons are studied both by
a modified variational analysis and by direct numerical
simulations. The effects of atoms feeding from the thermal cloud
are also considered. Finally, in the last section, the main
results of the paper are briefly discussed and summarized.

\section{The model}
Let us consider a BEC with two and three-body interactions
immersed in an optical lattice and an highly elongated   harmonic
trap. In the mean field approximation the system is described by
the following 3D GPE \cite{AGTF}
\begin{eqnarray}
i\hbar u_{t}&=&-\frac{\hbar^2}{2m}\nabla^{2}u + [\frac{m}{2}
(\omega_{\perp}^2 \rho^2 + \omega_{x}^{2}x^{2})+ V_{opt}(x)]u
\nonumber \\ && +\, g_{1}|u|^{2}u + g_{2}|u|^{4}u,
\end{eqnarray}
with $\rho$ denoting the radial distance, $g_{1}, g_{2}$ the
nonlinear coefficients corresponding to the  two and three-body
interactions, respectively,  $\omega_{\perp},  \omega_{x}$ the
radial and longitudinal frequencies  of the anisotropic trap
($\omega_{\perp} \gg \omega_{x}$) and $V_{opt}(x)=V_0 \sin^2(k x)$
the optical lattice applied only in the longitudinal direction.
The coupling constant of the two body interaction is related to
the s-wave scattering length $a_s$  and to the mass $m$ of the
atoms by the usual relation $g_1=4 \pi \hbar^2 a_s/m$. In view of
the strong anisotropy  of the trap we can  average the three
dimensional interaction over the radial density profile to reduce
the problem to an effective one dimensional one. More precisely,
we consider solutions of the form $u(r,t)
=\phi_{0}(\rho)\psi(x,t)$ where $\phi_{0} = \sqrt{\frac{1}{\pi
a_{\perp}^2}} \exp(-\frac{\rho^{2}}{2a_{\perp}^2})$ is the ground
state of the radial linear equation
\begin{equation}
-\frac{\hbar^2}{2m}\nabla_{\rho}^{2}\phi_{0} + \frac{m}{2}
\omega_{\perp}^{2}\rho^{2}\phi_{0} = \hbar
\omega_{\perp}\phi_{0}.
\end{equation}
Multiplying both sides of the GPE by $\phi_{0}$ and integrating
over the transverse variable, we obtain the following quasi 1D-GP
equation \cite{Zhang}
\begin{eqnarray}
i\hbar \psi_{t} =-\frac{\hbar^2}{2m}\psi_{xx} + (\frac{m}{2}
\omega_{x}^{2}x^{2} + V_{0}\sin^{2}(k x))\psi + \nonumber \\
\frac{g_{1}}{2\pi a_{\perp}^2}|\psi|^{2}\psi  + \frac{
g_{2}}{3\pi^2 a_{\perp}^4}|\psi|^{4}\psi.
\end{eqnarray}
For a discussion on the limits of applicability of this  1D
reduction see \cite{Brazh}. Introducing the dimensionless
variables
\begin{equation}
t = t\nu, x = kx,\;\; \nu = \frac{E_{R}}{\hbar},\;\; \alpha =
\frac{\omega_{x}^2}{4\nu^2},
\end{equation}
we reduce Eq. 3 to the following normalized 1D GPE equation with
cubic and quintic nonlinearities
\begin{equation}\label{gpe}
i\psi_{t} = -\psi_{xx} + g |\psi|^{2}\psi  + \chi |\psi|^{4}\psi +
\alpha x^{2}\psi - \varepsilon \cos(2x)\psi,
\end{equation}
where $a_{s} = a_{s0}g$, with $a_{s0}$ the constant scattering
length,  $g\in[-1,1]$ and $E_{R} = \hbar^2 k^2/2m$ is the recoil
energy of the lattice. In this normalization the wavefunction has
been scaled according to $\psi \rightarrow
\psi\sqrt{{2a_{s0}\omega_{\perp}}/{\nu}}$ and the parameters
$\varepsilon, \chi,$ are defined as $\varepsilon =
V_{0}/(2E_{R})$, $\chi = {g_{2}}\left(
{\nu}/{2a_{s0}\omega_{\perp}} \right)^{2}/{3\pi^2 \hbar \nu
a_{\perp}^4}$. In the following we will be mainly interested in
the case $g=0$ for which  it is convenient to  rescale the
wavefunction as $\psi \rightarrow (\frac{g_2}{3 \pi^2 a_{\perp}^4
E_R})^{1/4}\psi$. In this case the relation between the
dimensionless number of atoms $N$ and the physical one $N_p$ is
$N_p = (3\pi^{2} a_{\perp}^{4}E_{R}/(g_{2} k^{2}))^{1/2}N$ (values
of $N_p$ are typically in the range $ 4000 - 5000$ for $g_2/\hbar
\approx 10^{-26}cm^{6}s^{-1}, a_{\perp} = 2\mu$ m, $\lambda =
8.23\cdot 10^{-7}$m, $m=1.44\cdot 10^{-25}kg$).

Equation (\ref{gpe}) is obtained from the following Hamiltonian
\begin{equation}
H = \int_{-\infty}^{\infty} dx[|\psi_{x}|^2 + \frac{g}{2}|\psi|^4
+ \frac{\chi}{3}|\psi|^6  + V(x)|\psi|^2 ],
\end{equation}
as $i\psi_{t} = \delta H/\delta \psi^{\ast}$. This equation
appears also in the context of nonlinear photonic crystals
\cite{Gisin}. The relevance of the two-body interactions (cubic
term) for the formation of localized excitation of soliton type
has been largely investigated in the past decades in the field of
nonlinear optics and the existence of bistable solitons in optical
media with Kerr (cubic) and saturable (quintic) nonlinearity was
established for the case of a channel waveguide with step
potential \cite{Gisin}.

The cubic-quintic NLSE also appears as a model for the propagation
of optical pulses in double-doped optical fibers \cite{DeAngelis}
with an  effective refraction index given by $n = n_2 + n_4
|E|^2$, where $n_2 = n_{d_1} + n_{d_2}, n_4 = n_{d_2}$, and
$n_{d_i}$ are the contributions of dopants to the refraction index
(values and signs can be changed by proper choice of dopants).
Although Townes solitons have been presently found only in
connection with collapse of 2D elliptic shaped intensive beams in
homogeneous focusing Kerr media \cite{Fibich1}, we expect
gap-Townes solitons to be observed also in double-doped fibers
with Bragg gratings in the form of optical pulses and in photonic
crystal fibers in the form of spatial optical solitons.

Finally, we remark that a pure quintic NLS equation  with
repulsive interactions also appears in connection with a
Tonks-Girardeau gas in the local density approximation \cite{Kol}.
It this context the field equation is
\begin{equation} i\hbar{\phi_{t}} =
-\frac{\hbar^2}{2m}\phi_{xx} + V(x)\phi  +
\frac{\pi^{2}\hbar^{2}}{2m}|\phi|^{4}\phi,
\end{equation}
also written  in normalized form as
\begin{equation}
iu_{t} + u_{xx} -\alpha x^2 u + \epsilon\cos(2x)u -|u|^4 u = 0,
\end{equation}
which corresponds to the case  $g = 0, \chi = 1$ in
Eq.(\ref{gpe}). Here normalization has been made by introducing
dimensionless variables $x = kx, t = t\nu, u = \phi\sqrt{\pi/k},$
with $\int dx |\phi|^2=N_p=\frac {N}{\pi}$. In the next sections
we use the variational approach \cite{mal}, the self consistent
method \cite{salerno} and direct numerical integrations,  to study
localized states of the quintic NLS in presence of an optical
lattice  for both attractive and repulsive interactions.

\begin{figure}[ht]
\centerline{
\includegraphics[width=8.cm,height=6.cm,angle=0,clip]{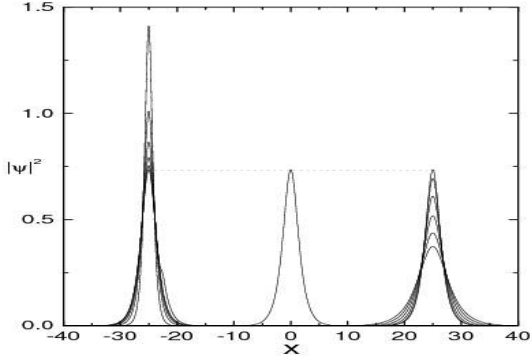}}
\caption{Early stages of the evolution toward collapse (left
curves) or decay (right curves) of the critical soliton (central
curve) in absence of external potentials ($\alpha=0,
\varepsilon=0$). The critical soliton in the center refers to the
value $\chi=-1$ and $N=2.7207$, while the collapse and the decay
evolution refer to $\chi=-1.01$ and $\chi=-0.99$, respectively,
with all other parameters unchanged. Snapshots were taken at
regular time intervals of $5.2$ for collapse and of $10$ for
decay. Plotted quantities are in normalized units.}\label{fig1}
\end{figure}
\begin{figure}\centerline{
\includegraphics[width=4.cm,height=4.cm,angle=0,clip]{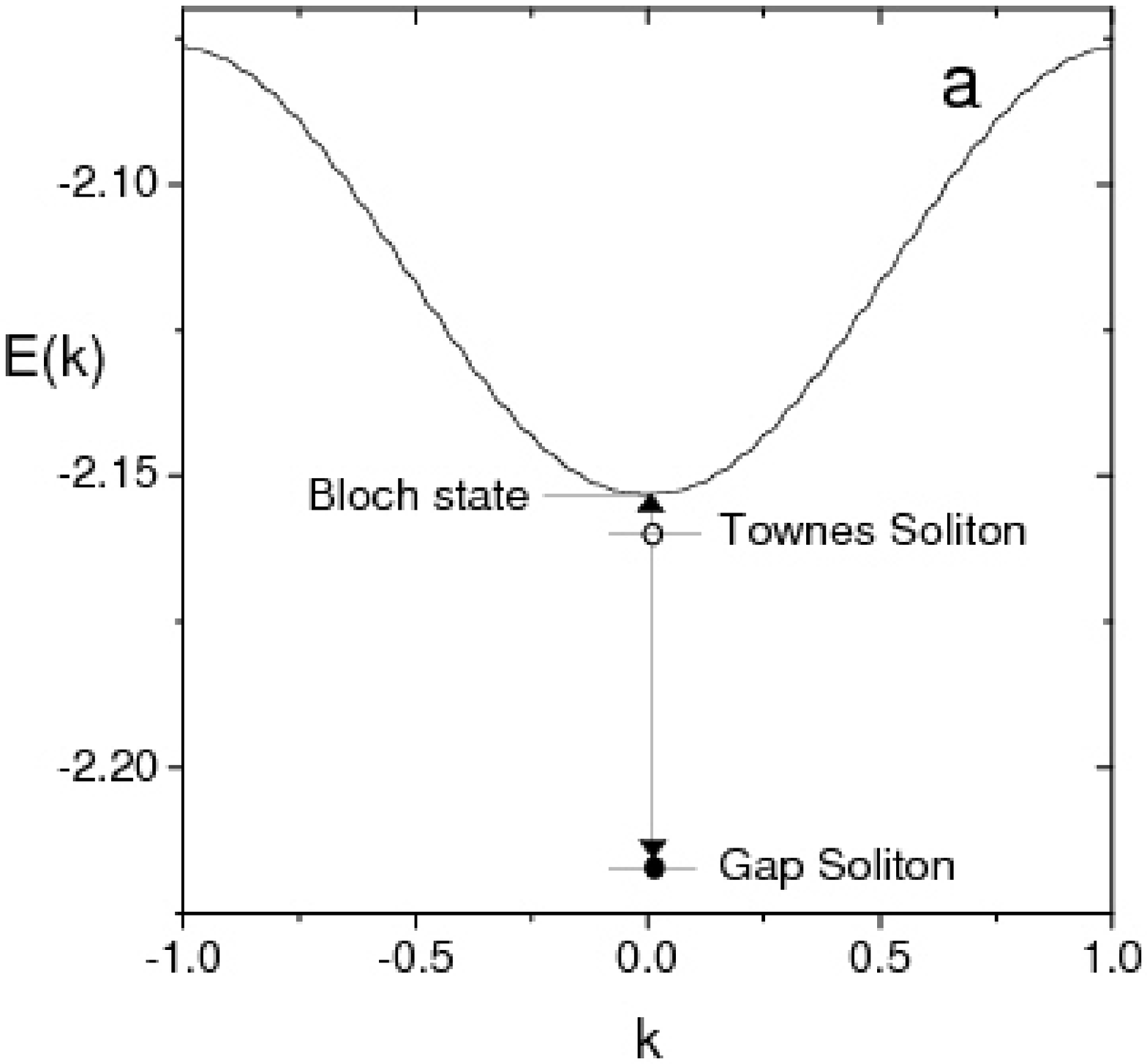}
\includegraphics[width=4.cm,height=4.cm,angle=0,clip]{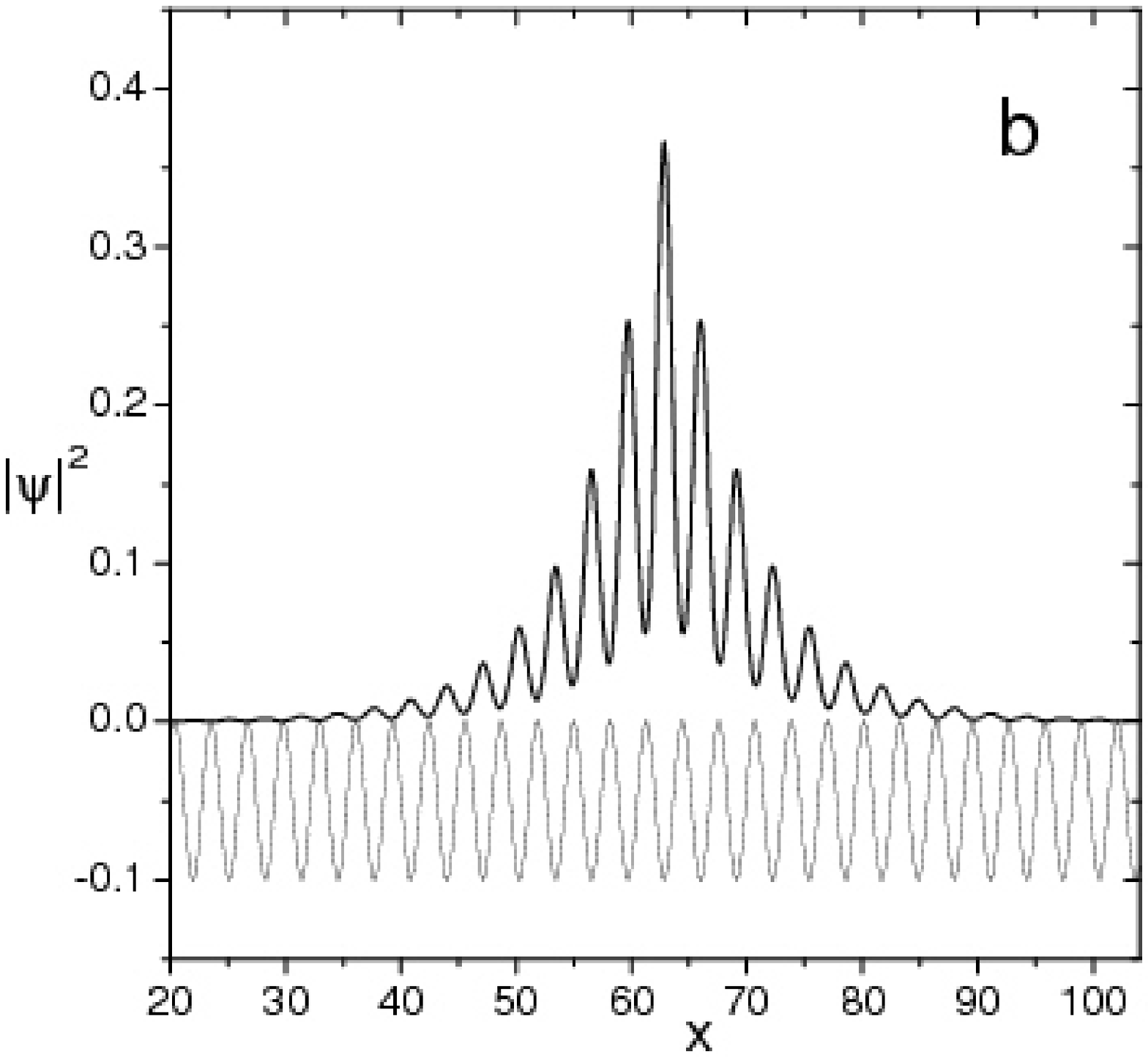}
} \centerline{
\includegraphics[width=4.cm,height=4.cm,angle=0,clip]{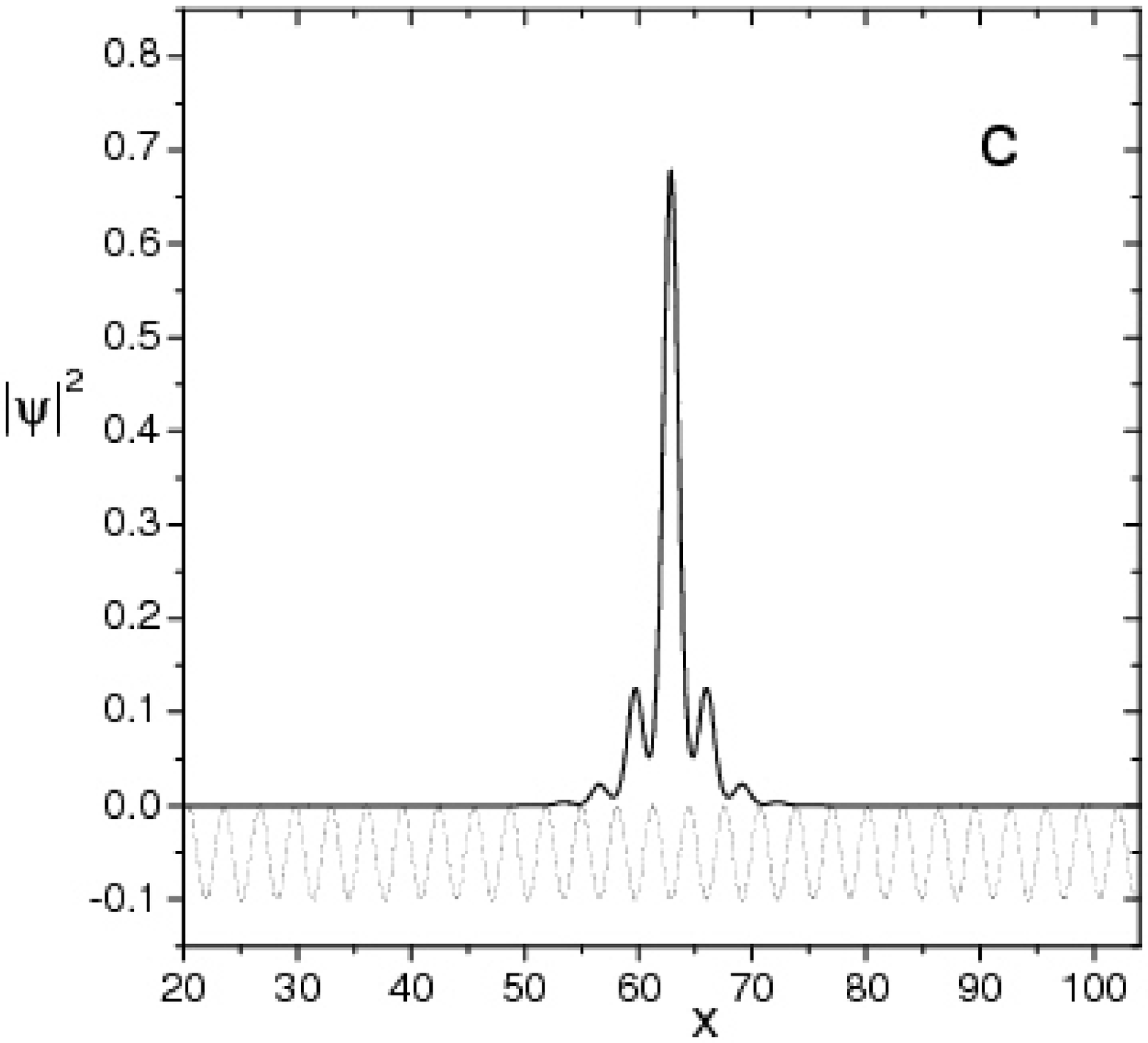}
\includegraphics[width=4.cm,height=4.cm,angle=0,clip]{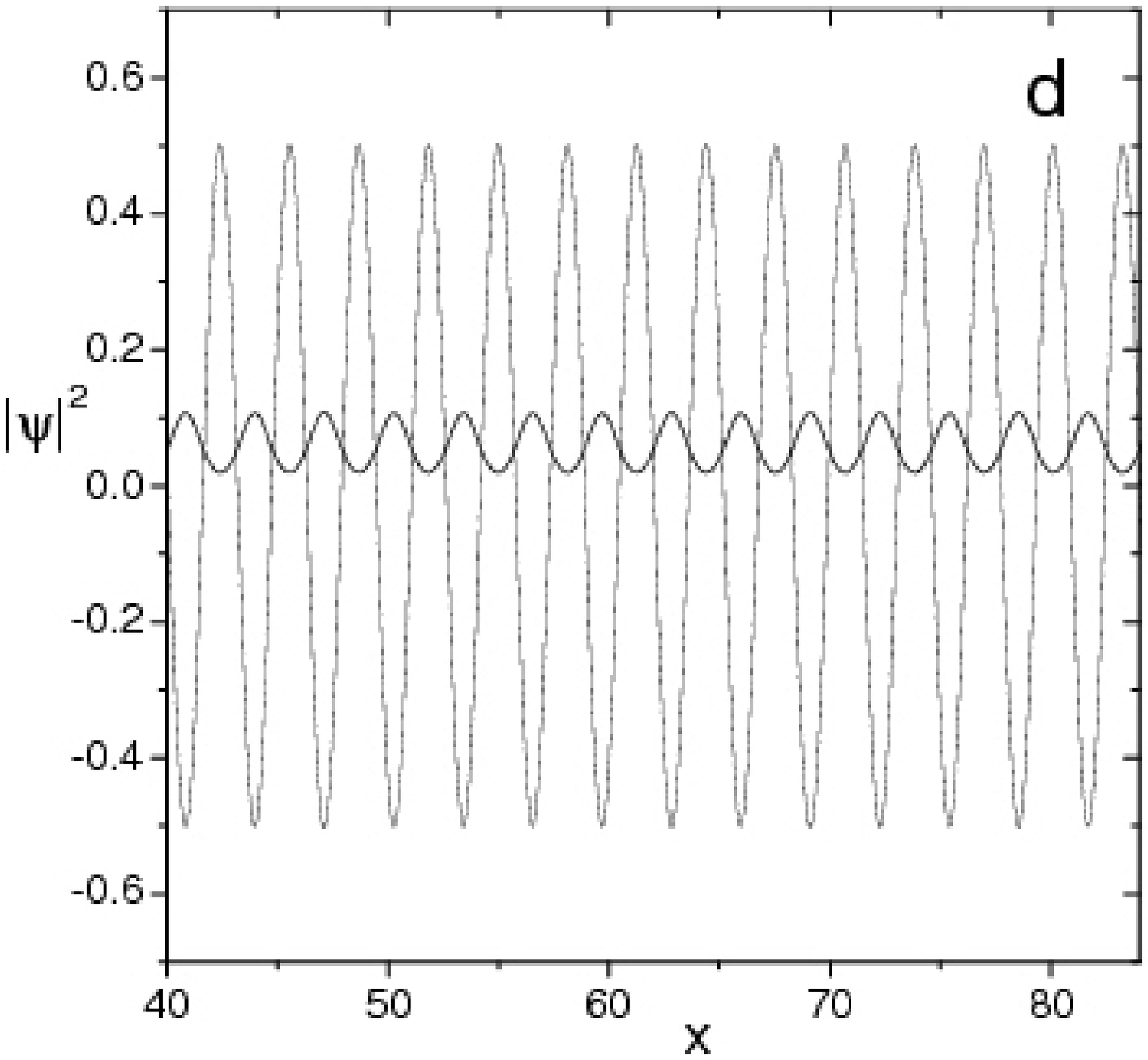}
} \caption{{\bf Panel a}. Lower band  and localized bound states
of the quintic NLS with attractive interactions for parameter
values $\chi=-1$, $\varepsilon=5$. The arrows show the decay
pattern of the Townes bound state (open circle) for $N<N_{cr}$
(upper arrow) and $N>N_{cr}$ (lower arrow). {\bf Panel b}. Townes
soliton (thick line) corresponding to the unstable bound state
level (open circle) in panel a, with $N=N_{cr}=0.4616$ and
$E=-2.1576$. {\bf Panel c}. Gap soliton (thick line and filled
circle in panel a) obtained from the unstable Townes state for a
slightly above critical value of $N$ ($N=0.4657, E=-2.2177$). {\bf
Panel d}. The Bloch state (thick line) at the bottom of the band
into which the Townes soliton decays for a slightly sub-critical
value of $N$ ($N=0.4586, E=-2.1531$). The OL is depicted as a thin
dotted line scaled by a factor 100 and shifted down by $.05$ in
panel b,c, and scaled by a factor $10$ in panel d. Plotted
quantities are in normalized units.} \label{fig2}
\end{figure}
\begin{figure}\centerline{
\includegraphics[width=2.95cm,height=7.6cm,angle=0,clip]{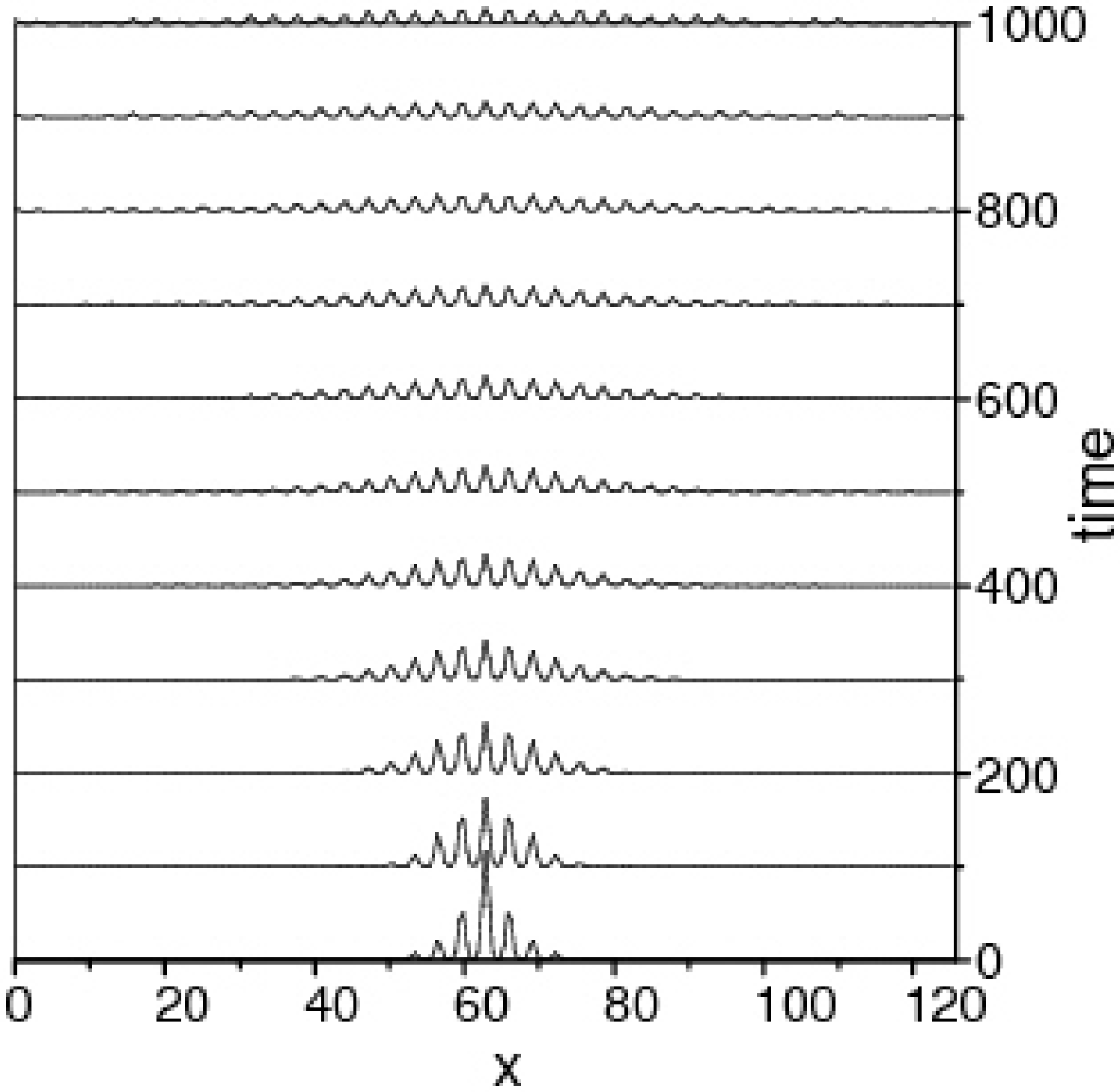}
\includegraphics[width=2.95cm,height=7.9cm,angle=0,clip]{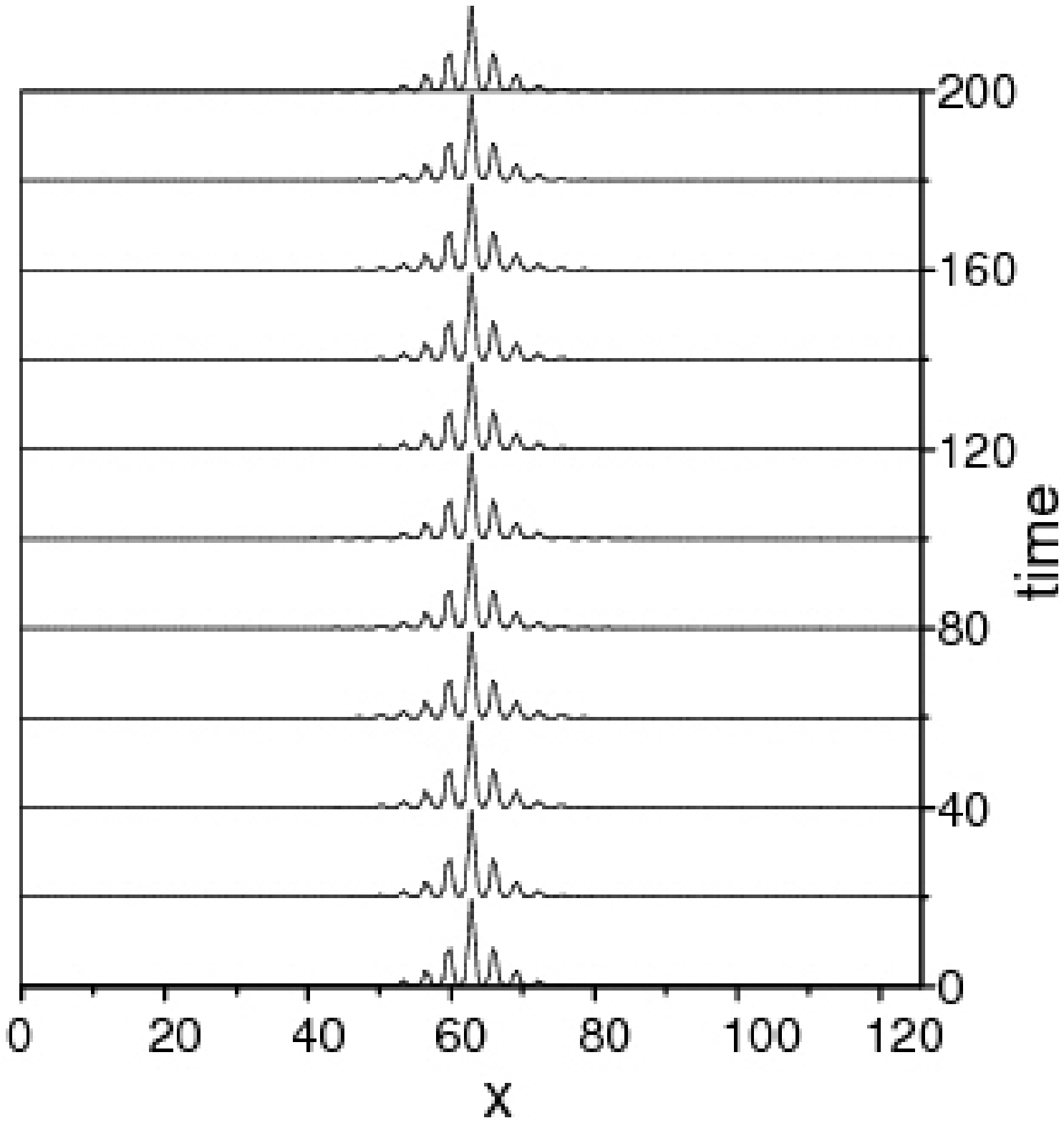}
\includegraphics[width=2.95cm,height=8.5cm,angle=0,clip]{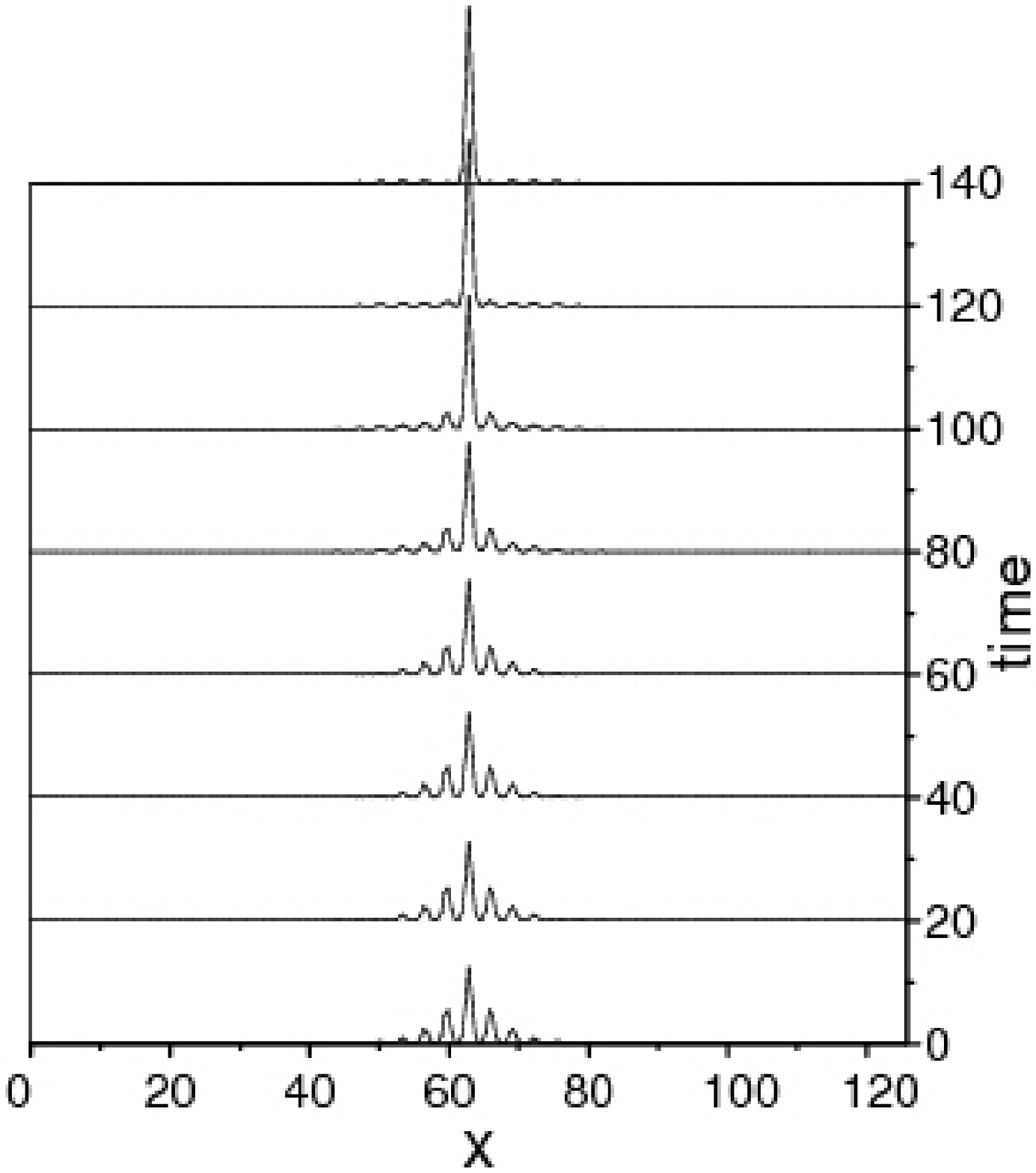}}
\caption{Time evolution of the Townes soliton in Fig.\ref{fig2}
(central panel) compared with the one obtained from numerical
integration of the quintic NLS with two slightly different values
of $N$:  $N=1.1 N_{cr}$ (right panel) and $N=0.9 N_{cr}$ (left
panel). Parameters are fixed as in Fig. \ref{fig2} and plotted
quantities are in normalized units.} \label{fig3}
\end{figure}

\section{Attractive interactions}
To obtain analytical predictions for the existence of stable
localized solutions of Eq.(\ref{gpe}) with $\alpha=0$ and for
attractive interactions ($g,\chi<0$ in Eq.(\ref{gpe})), we use the
variational approximation (VA) with a Gaussian ansatz for the
fundamental soliton
\begin{equation}
\psi(x,t)=A\exp \left( -i\mu t- \frac12 a x^{2}\right) ,
\label{ansatz}
\end{equation}
with $\mu$ denoting the chemical potential and $A$, $a$, the
amplitude and the square root of the reciprocal width of the
soliton, respectively. Following the standard VA (see \cite{Mal}
for a review) we derive the following effective Lagrangian
\begin{equation}
\bar{L} = \frac{A^{2}\sqrt{\pi}}{2\sqrt{a}}\left( \mu - \frac{a}{2}
 + \varepsilon e^{-1/a} - \frac{g A^{2}}{2\sqrt{2}} -
\frac{\chi A^{4} }{3\sqrt{3}}\right),
\end{equation}
from which the stationary equations $\partial \bar{L}/\partial
a=\partial \bar{L}/\partial A=0$ for parameters $a$ and $A$ are
obtained as
\begin{eqnarray}
&& \mu = -\frac{a}{2} (1 - \frac{2 N^2
\chi}{3^{3/2}\pi})+\varepsilon (\frac{2}{a}-1) e^{-1/a}
+\frac{gN\sqrt{a}}{2\sqrt{2\pi}}, \label{var1} \cr && \frac{2 N^2
\chi}{3^{3/2}\pi}  + \frac{g N}{2\sqrt{2\pi a}}=
\frac{2\varepsilon}{a^2} e^{-1/a} -1.\label{var2}
\end{eqnarray}
Here the number of atoms has been expressed in terms of $a, A,$ as
$N=A^{2} \sqrt \frac\pi a$. Note that for $\varepsilon =0, g=0$,
Eqs. (\ref{var2}) predicts a maximum value of $N$  for the
existence of soliton given by $N_{cr}= \sqrt{3^{3/2}\pi/(2
\chi)}$. This value should  be compared with the critical norm
$N_{cr}^{(e)} =(\pi/2)\sqrt{3/|\chi|}$ obtained for the exact
Townes soliton  of Eq. (\ref{gpe}) with $\varepsilon =0, g=0,
\alpha =0$, \cite{Gaididei}
\begin{equation}
\psi_s = \exp(i\mu t)\psi,\;\;\;  \psi
=\frac{(3\mu/|\chi|)^{1/4}}{\sqrt{\mbox{cosh}(2 \sqrt{\mu}x)}}.
\label{townes}
\end{equation}
We see that the VA value deviates from the exact one by only
$5\%$. Notice that the above solution has zero energy and is
marginally stable ($dN/d\mu=0$) so it exists only for the single
value of the norm $N=N_{cr}^{(e)}$. Since for $N> N_{cr}^{(e)}$
the solution collapses in a finite time, while for $N <
N_{cr}^{(e)}$ it decays into the uniform background, it appears as
the analogue of the Townes soliton of the cubic 2D NLS equation
with attractive interactions. In Fig.\ref{fig1} we depict the
early stages of the time evolution toward collapse (decay) for an
initial norm which is slightly increased (decreased) with respect
to $N_{cr}^{(e)}$.

For $\varepsilon \neq 0$ Eq. (\ref{var2}) shows that the number of
particles attains a minimum value $N_{{\rm thr}}=\left[ 6 \pi
\sqrt{3}( 1- 8 \varepsilon e^{-2})\right]^{1/2}$ at $a=1/2$, i.e.
there exists a threshold value of the norm which is necessary to
create a soliton. The expression for $N_{{\rm thr}}$ shows that
the threshold exists for a relatively weak lattice and disappears
if $\varepsilon $ exceeds the value $\varepsilon
_{0}=e^{2}/8\approx \allowbreak 0.\,\allowbreak 9236$. From this
analysis it is clear that the optical lattice is quite effective
to stabilize solitons against decay when  $N$ is in the interval
$N_{{\rm thr} }< N < N_{{\rm cr}}$. Collapse and decaying into
extended states is expected for $N > N_{{\rm cr}}$ and $N <
N_{{\rm thr}}$, respectively. The VA makes it possible to predict
stability of the solitons on the basis of the {\it
Vakhitov-Kolokolov} (VK) criterion \cite{Berge}, according to
which  a necessary condition for stability is given by $d\mu
/dN<0$. Notice that the above variational equations are very
similar to the ones derived in Ref. \cite{bms03} for the 2D cubic
NLS (with exactly the same value of $\varepsilon _{0}$), this
being a further confirmation of the analogy between the 1D quintic
NLS and the usual 2D GPE. We remark that, since $N_{cr}$ does not
dependent on $\varepsilon$, the prediction of the VA is that the
optical lattice does not affect collapse. We shall check these
results in the next section.
\begin{figure}\centerline{
\includegraphics[width=4.2cm,height=4.5cm,angle=0,clip]{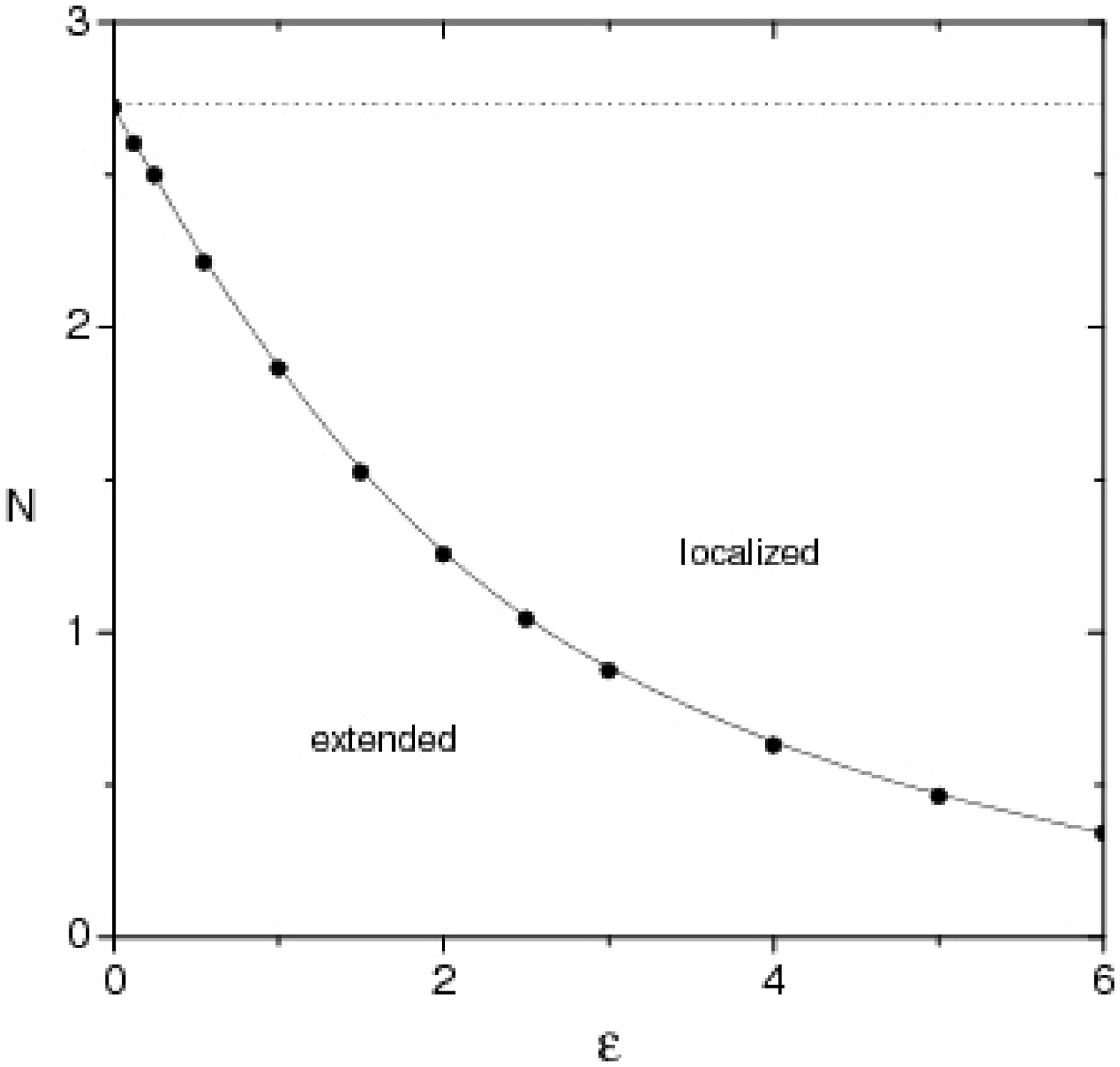}
\includegraphics[width=4.2cm,height=4.5cm,angle=0,clip]{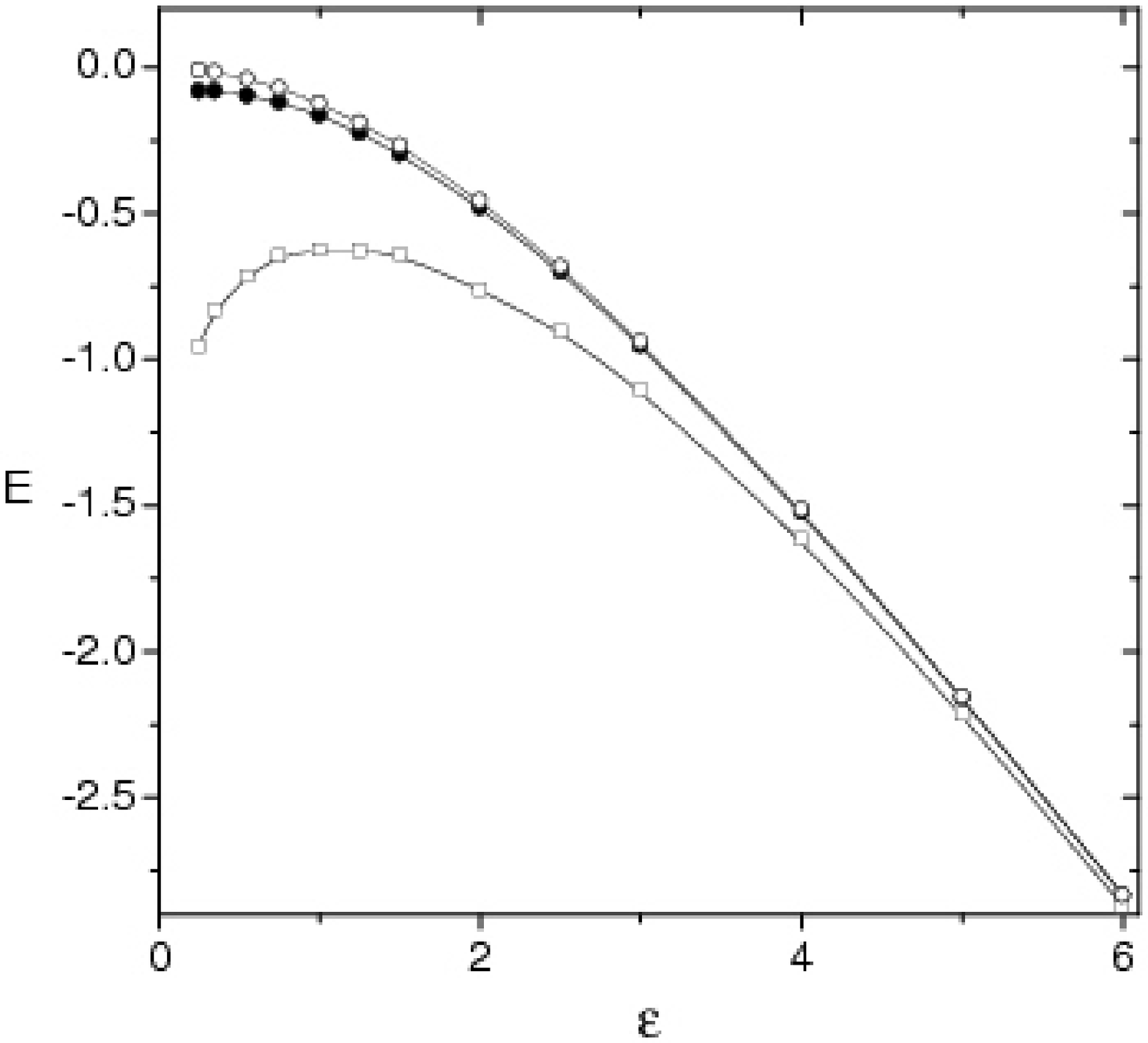}}
\caption{ Left Panel. Delocalizing transition threshold obtained
by direct numerical integrations (filled dots) of the quintic NLS
with $\chi=-1.0$. The horizontal dotted line corresponds to the
critical threshold for collapse in absence of the OL:
$N_{cr}=\frac \pi 2 \sqrt{\frac3 \chi}$. Right Panel. Energies
versus $\varepsilon$ of the gap soliton (bottom curve with open
squares), Townes soliton (middle curve with filled circles) and
Bloch state at the bottom of the band (top curve with open
circles) for the quintic NLS with $\chi=-1.0$. Plotted quantities
are in normalized unit. } \label{fig4}
\end{figure}

\section{Gap-Townes solitons and arresting collapse in
1D BEC in OL} The existence of the exact  Townes soliton
(\ref{townes}) of the quintic NLS with $\varepsilon=0$ makes
natural to ask whether this solution could exist also in presence
of the OL. In view of the analogies between 1D quintic NLS and  2D
GPE, the existence of this solution for $\varepsilon \neq 0$ could
shine some light on the role of the OL in controlling collapse, as
well as, on the origin of the delocalizing transition observed in
these systems \cite{BS}. In the following we use a self consistent
approach \cite{salerno} to determine the critical value of $N$ for
the unstable solution to exist.

In Fig.\ref{fig2}a we report the band structure and the localized
states obtained with the self-consistent approach for the case
$g=0, \alpha = 0$ in Eq. (\ref{gpe}). We see that there are two
bound state levels just below the band. The level closer to the
band edge (open circle) corresponds to an unstable localized state
while the other corresponds to a stable gap soliton. The
corresponding wavefunctions are depicted in Figs. \ref{fig2}b,c,
respectively. We find that for a fixed value of $\varepsilon$
there is only one value $N$ for which the unstable state exists
and its behavior resembles the one of the Townes soliton depicted
in Fig. \ref{fig1}. In this case, however, we have that for a
slightly overcritical  value of N the unstable soliton starts to
shrink and its energy decreases as for collapsing solutions until
it reaches the gap soliton level below, at which the shrinking
stops. The transition from the unstable Townes soliton to the
stable gap soliton is represented in Fig.\ref{fig2}a by the lower
arrow. If the norm is slightly decreased below $N_{cr}$ the
unstable state completely delocalizes into the Bloch state at the
bottom of the band shown in Fig.\ref{fig2}d (this decay is
represented in Fig.\ref{fig2}a by the upper arrow). Due to this
behavior we refer to the unstable state as {\it gap-Townes
soliton}, this name being  also justified by the fact that for
$\varepsilon \rightarrow 0$ the critical norm reduces to
$N_{cr}^e$ (see left panel of Fig. \ref{fig4}). The decaying
property of the gap-Townes soliton is illustrated in
Fig.\ref{fig3} where the time evolution obtained from direct
numerical integrations of Eq. \ref{gpe} (with $g=0, \alpha =0$) is
shown. The left  (right) panel of this figure shows the decay into
the extended state (gap soliton) when the initial norm is slightly
decreased (increased) with respect to $N_{cr}$. Notice that the
transition into the gap soliton state is much more rapid than the
decay into the extended state, this being a reminiscence of the
collapse occurring at $\varepsilon=0$. Since the energy of a
slightly overcritical gap-Townes  solution should go to $-\infty$,
as for any collapsing solution,  we see that the existence of a
stable localized state which lies in energy below the unstable
state is indeed a mechanism for arresting collapse in the system.
\begin{figure}\centerline{
\includegraphics[width=6.cm,height=6.cm,angle=0,clip]{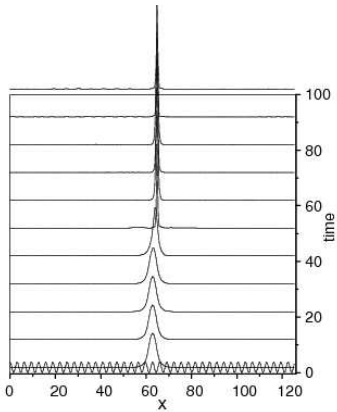}}
\caption{Time evolution of an intersite symmetric intrinsic
localized mode for the attractive case $\chi=-1$ with $N=2.847$
and $\epsilon=0.1$. Notice that collapse starts  after the mode
has decayed into an onsite symmetric one (ground state). The
periodic potential is shown together with the initial state at
$t=0$. Plotted quantities are in normalized unit.} \label{fig5}
\end{figure}
Also note that a gap-Townes soliton appears as a separatrix
between localized and extended states. The fact that its energy
lies below the lower band edge implies the existence of a
delocalizing threshold in the number of atoms below which
localized state of the quintic GPE with OL cannot exist, this
being a remarkable difference between cubic and quintic
nonlinearities (in the cubic case  the threshold exists only in
the higher dimensional case \cite{BS}). In Fig. \ref{fig4}a the
threshold curve separating localized and extended states in the
$(N,\varepsilon)$ plane is depicted (this curve coincides with the
norm of the gap-Townes soliton as a function of $\varepsilon$).
Notice that for $\varepsilon=0$ the norm coincides with  the exact
value $N_{cr}^{(e)}$ of the solution in (\ref{townes}), suggesting
that the family of gap-Townes soliton for $\varepsilon \ne 0$
originates from the Townes soliton state at $\varepsilon=0$.

Numerical investigations show that collapse occurs when the
strength of the optical lattice becomes small and the number of
atoms overcomes the critical threshold for Townes soliton at
$\varepsilon=0$. For larger values of $N$ the width of the soliton
becomes small compared to the period of the periodic potential and
the OL becomes ineffective  for arresting collapse. We find that
solitons extended on several potential wells are the ones which
are better stabilized by the optical lattice against collapse. An
indication of occurrence of collapse at small values of
$\varepsilon$ is obtained from the right panel of Fig. \ref{fig4}
in which we depict the energies of the Bloch state at the bottom
of the band, of the gap-Townes soliton immediately below the band
and of the gap soliton,  are reported as a function of
$\varepsilon$. We see that while the energies of the Bloch state
and of the gap- Townes soliton monotonically increase with
decreasing $\varepsilon$, the energy of the gap soliton reaches a
maximum around $\varepsilon\approx 1$ and then starts to decrease
as $\varepsilon \rightarrow 0$. This behavior is consistent with
the fact that a collapsing solutions with energy equal to
$-\infty$ should exist at $\varepsilon=0$. A study of collapse for
small $\varepsilon$ is numerically difficult to perform and
requires more investigations. It is interesting to note, however,
that besides stable gap-soliton the OL allows the existence of
metastable states which are symmetric around a maximum of the
potential (instead than a minimum as for usual gap-solitons).
These intersite symmetric states have higher energy than the
onsite symmetric ones and decay into the ground state after some
time. Such states can be used to delay the time of collapse for
$N> N_{{\rm cr}}$. Indeed, the collapse of these metastable states
begins when their instability sets in, as one can see from Fig.
\ref{fig5} (notice that the matter moves first into a single
potential well and then starts to collapse).

\section{Repulsive interactions}
To investigate the existence of localized states in the repulsive
quintic NLS with periodic potential we apply first a variational
analysis based on the following ansatz for the soliton waveform
\cite{BS}
\begin{equation}
\psi(x,t)= A \frac{\sin(a x)}{a x}\exp \left( -i\mu t\right).
\label{ansatz1}
\end{equation}
Performing the same analysis as before, we get the averaged
Lagrangian
\begin{equation}
\bar{L} = \frac{A^{2}\pi}{2a}\left( \mu  - \frac{a^2}{3}  +
\varepsilon (1 - \frac{1}{a})
 - \frac{11\chi A^{4}}{60} - \frac{gA^{2}}{3} \right)
\end{equation}
from which the following equations for the soliton parameters are
obtained
\begin{eqnarray}
&& \mu=\varepsilon (\frac{2}{a}-1) - \frac{1}{3} a^2 +
\frac{11\chi N^2 a^2}{60\pi^2} +
\frac{gNa}{3\pi},
\label{var1b} \cr && \frac{N^2 \chi}{ \pi^2} + \frac{10gN}{11\pi a}
=\frac{20}{11}(\frac32\frac{\varepsilon}{a^3}-1).\label{var2b}
\end{eqnarray}
Using the fact that the norm and the soliton parameters  are
related by $N=\pi A^2/a$, one can show that
\begin{equation} \label{aeps}
  A = \Bigl[ \frac{30 N^3 \varepsilon}{2\pi(11 N^2 |\chi|+20\pi^2)}
  \Bigr]^{1/6},
\end{equation}
i.e. the VA predicts the existence of a single set of soliton
parameters $a, A$ for a given norm $N$ and strength $\varepsilon$
of the OL. This is similar  to what obtained in Ref. \cite{BS} for
the localized solutions of the repulsive 2D GPE with OL.
\begin{figure}\centerline{
\includegraphics[width=4.cm,height=4.cm,angle=0,clip]{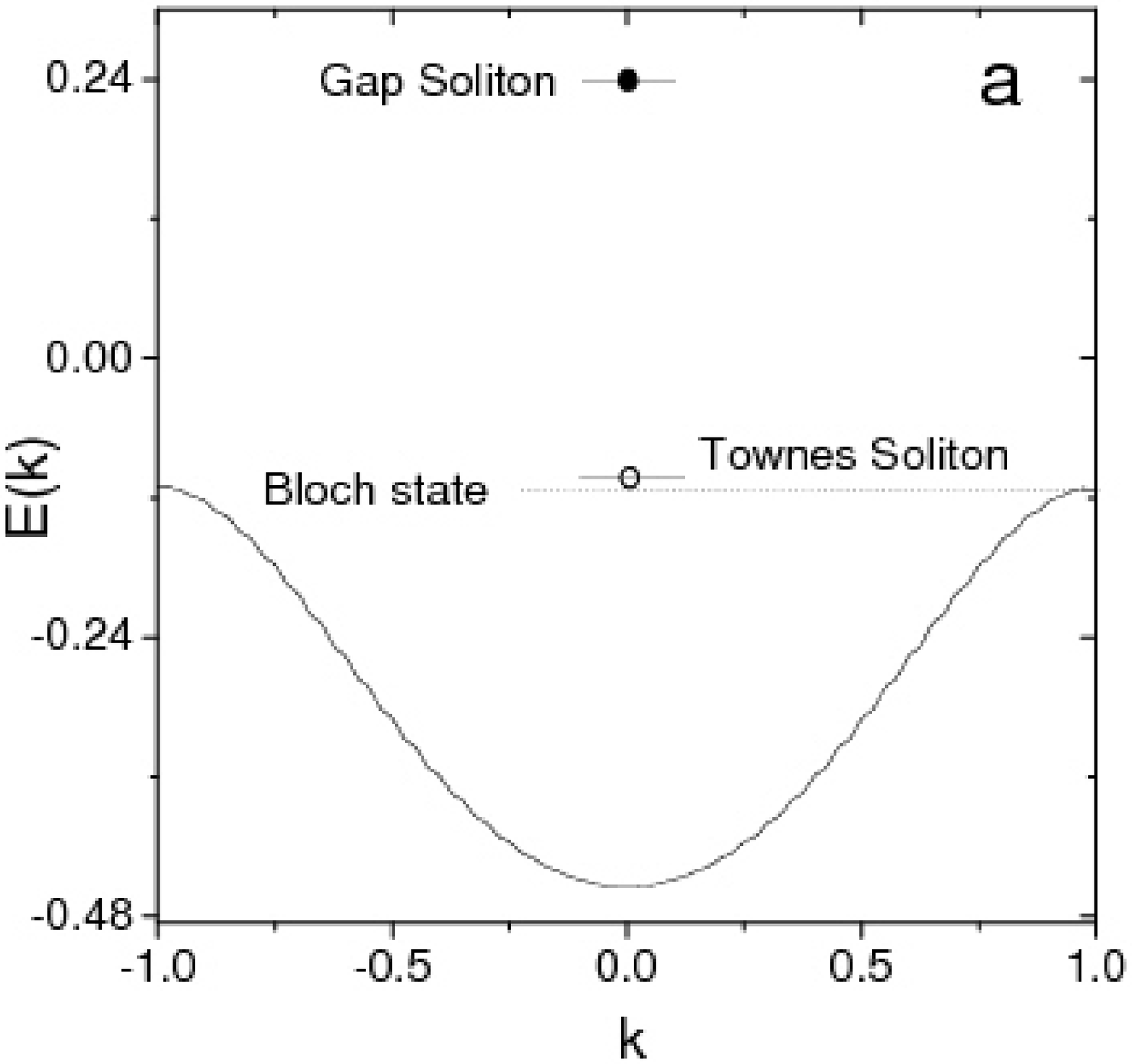}
\includegraphics[width=4.cm,height=4.cm,angle=0,clip]{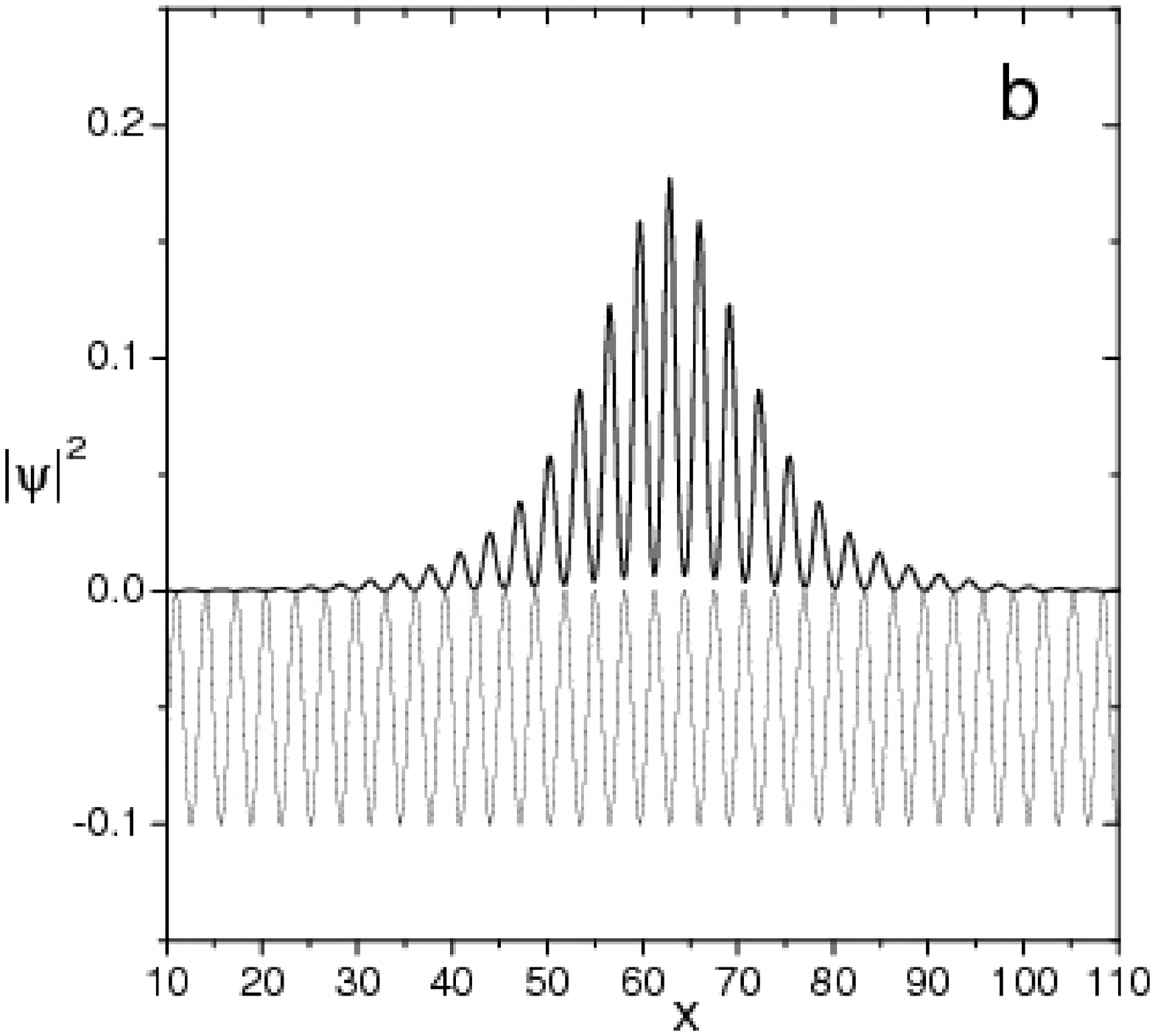}
} \centerline{
\includegraphics[width=4.cm,height=4.cm,angle=0,clip]{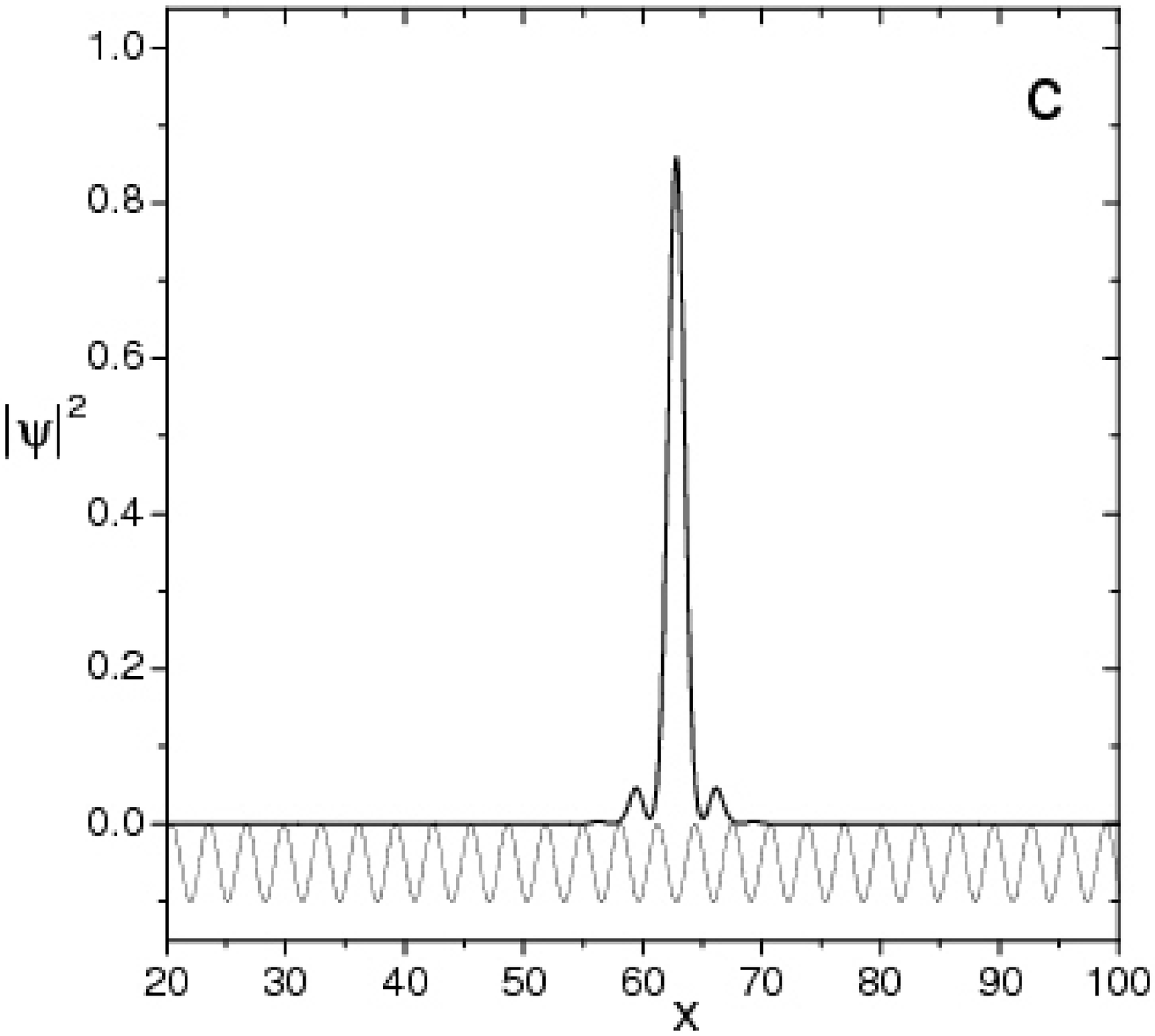}
\includegraphics[width=4.cm,height=4.cm,angle=0,clip]{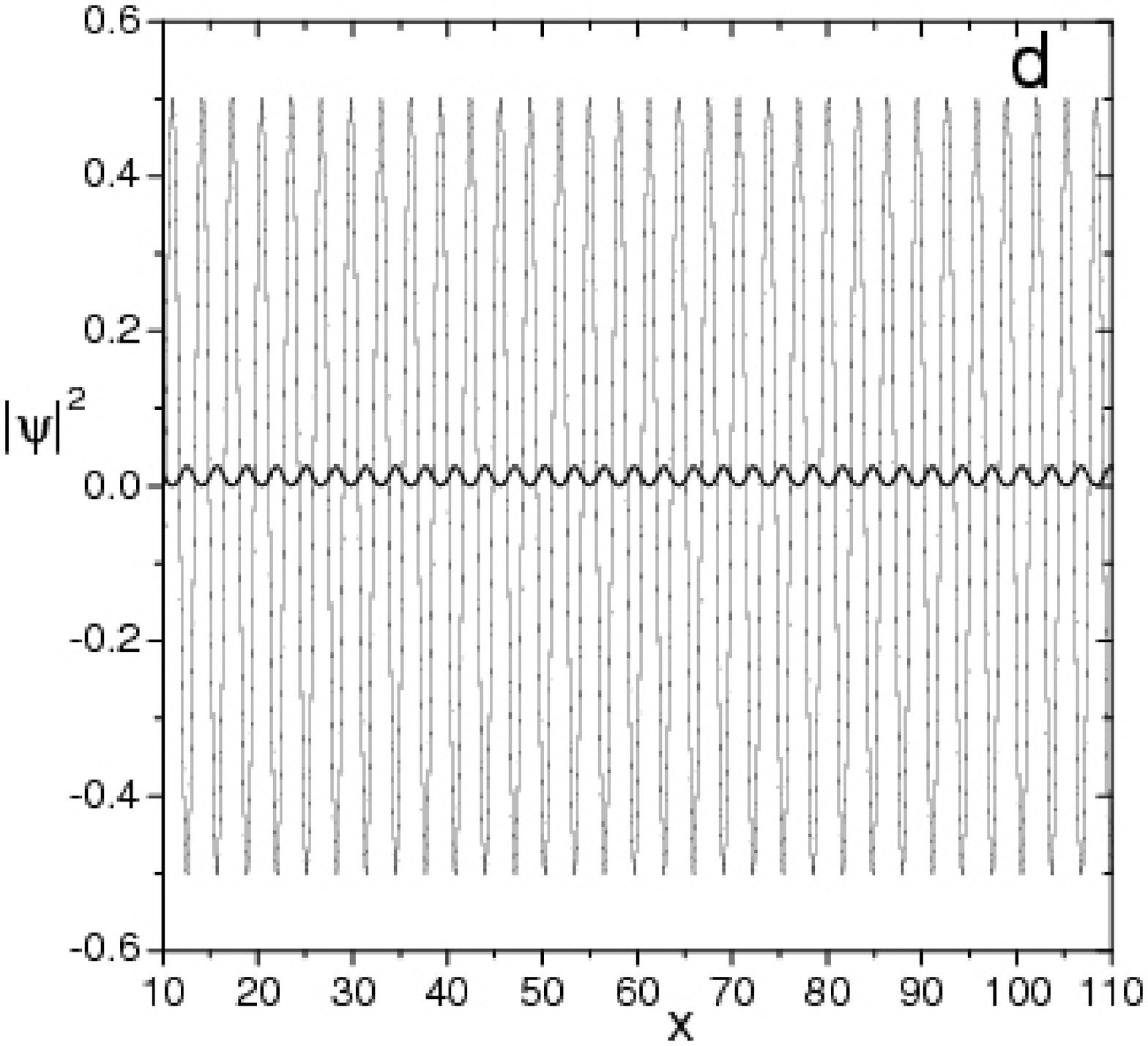}
} \caption{Panel a. Lower band  and localized bound states of the
quintic NLS with repulsive interactions for parameter values
$\chi=1$, $\varepsilon=2.0$. The arrows show the decay pattern of
the Townes bound state (open circle) for $N<N_{cr}$ (upper arrow)
and $N>N_{cr}$ (lower arrow). Panel b. Townes soliton (thick line)
corresponding to the unstable bound state level (open circle) in
panel a, with $N=N_{cr}=1.5565$ and $E=-0.1047$. Panel c. Gap
soliton (thick line and filled circle in panel a) obtained from
the unstable Townes state for a slightly above critical value of
$N$ ($N=1.5965, E=0.2537$). Panel d. The Bloch state (thick line)
at the bottom of the band into which the Townes soliton decays for
a slightly sub-critical value of $N$ ($N=1.5065, E=-0.1097$). The
OL is depicted as a thin dotted line scaled by a factor 20 and
shifted down by $.05$ in panel b,c, and scaled by a factor $10$ in
panel d. Plotted quantities are in normalized units.} \label{fig6}
\end{figure}
In the case of a Tonks-Girardeau gas in OL these parameters can be
estimated as follows . From Eq.(\ref{aeps}) we get that $A \approx
(30 N\varepsilon/22 \pi)^{1/6}$ and from the expression of $N$ we
get  for the soliton width $w = 1/a  \approx (22 N^2/(30\pi^2
\varepsilon))^{1/3}.$ For a number of particles $N=314$
(corresponding in physical units to $N_p = 100$) and for $V_0 = 4
E_{R}$,so $ \varepsilon = 2$, the width of the gap soliton is
estimated to be $w \approx 15.5$ corresponding to $3-4$ lattice
periods.

\begin{figure}\centerline{
\includegraphics[width=3.cm,height=6.5cm,angle=0,clip]{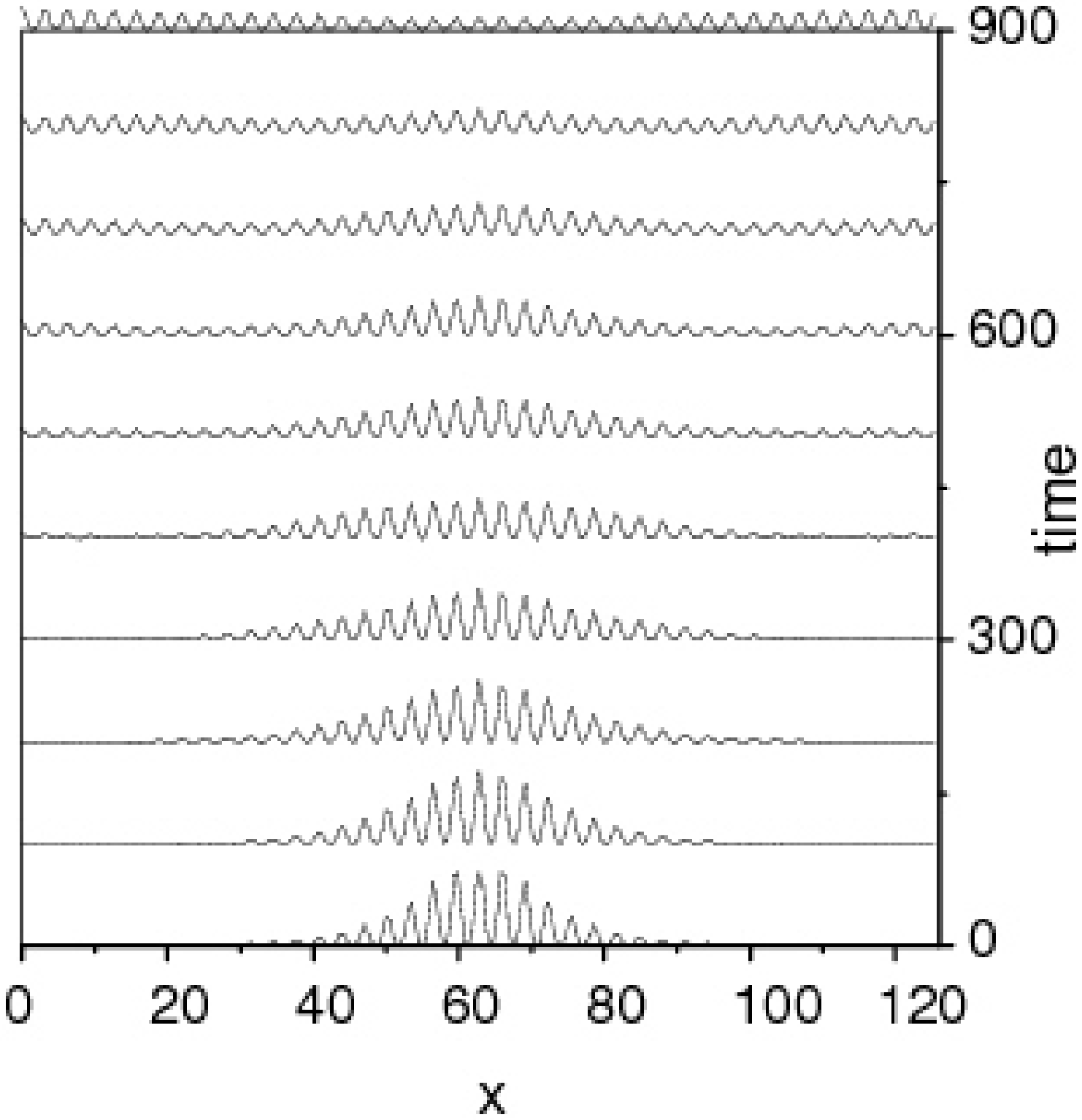}
\includegraphics[width=3.cm,height=6.8cm,angle=0,clip]{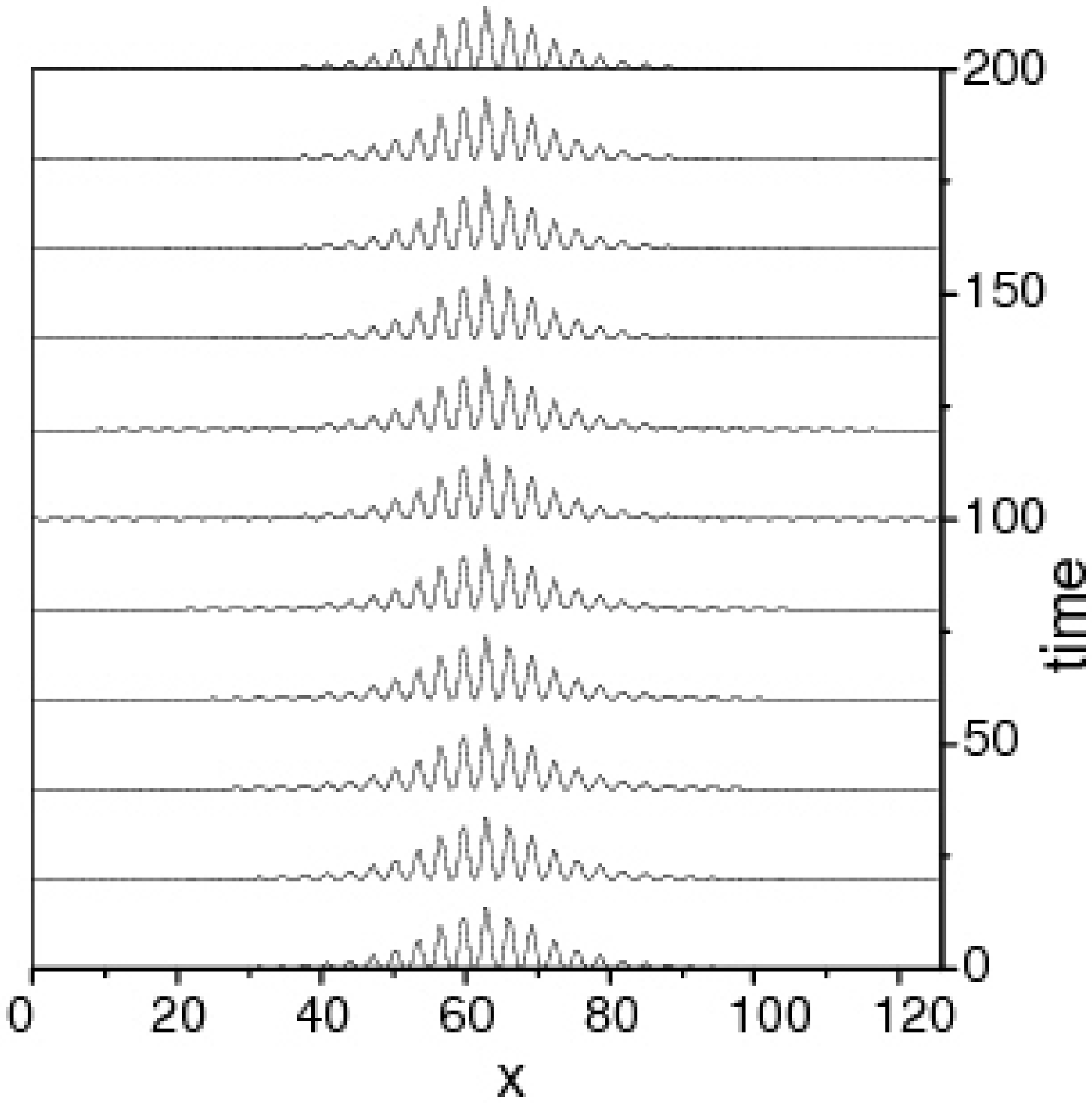}
\includegraphics[width=3.cm,height=7.5cm,angle=0,clip]{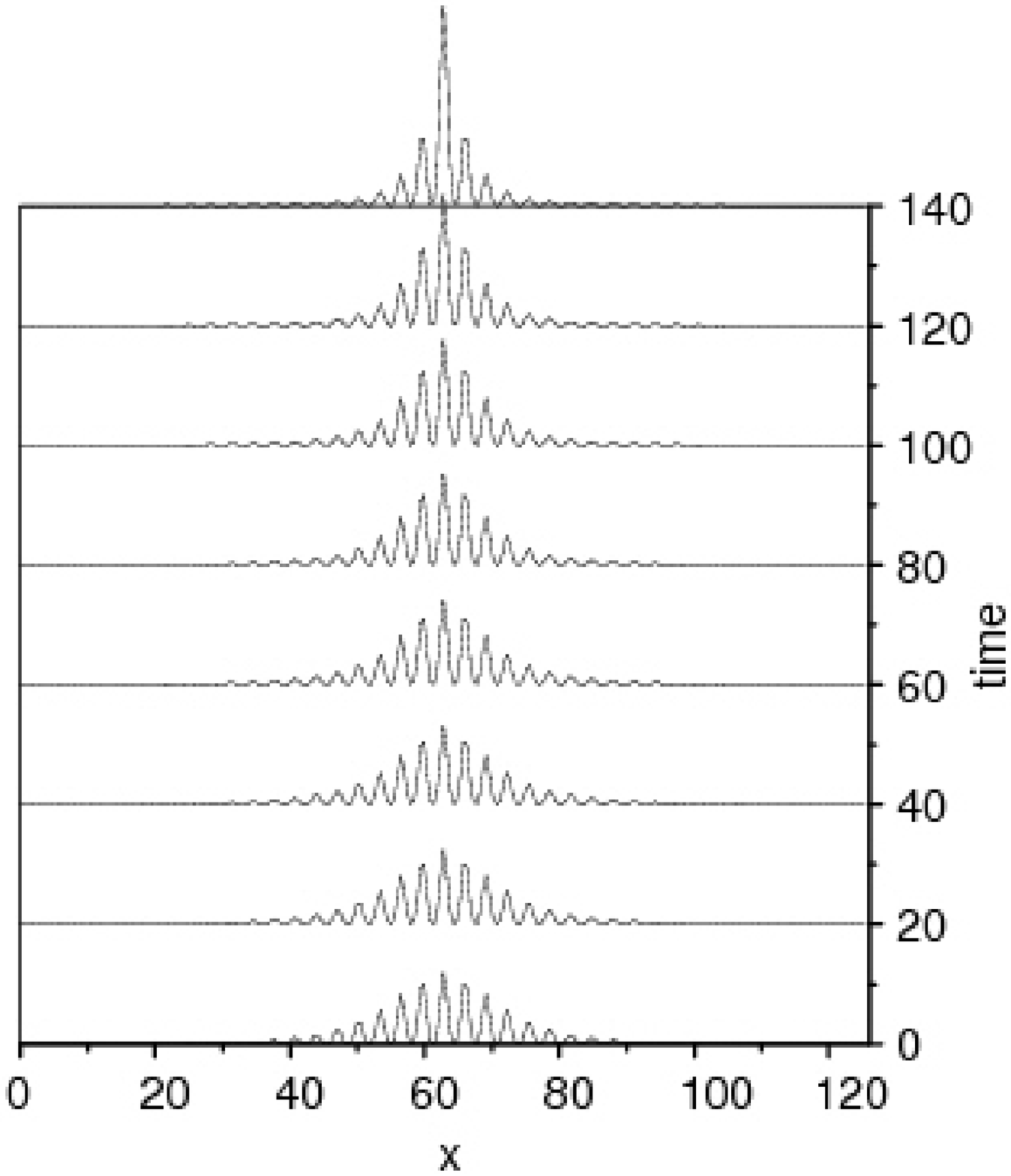}}
\caption{Time evolution of the Townes soliton in Fig.\ref{fig6} as
obtained from numerical integration of the quintic NLS with
repulsive interaction and  three slightly different values of
number of atoms: $N_{cr}=1.5565$ (central panel), $N=1.05 N_{cr}$
(right panel) and  $N=0.95 N_{cr}$ (left panel). Other parameters
are fixed as in Fig.\ref{fig6}. Plotted quantities are in
normalized units.} \label{fig7}
\end{figure}

The band structure associated to gap solitons of the repulsive
quintic NLS equation has been  investigated by means of the
self-consistent method. In Fig. \ref{fig6}a we show the lower band
and the localized bound states for the case $g=0, \alpha =0$ in
Eq. (\ref{gpe}). We see that above the band edge there are two
bound state levels, one immediately above (open circle)
corresponding to an unstable gap-Townes soliton, the other more
separated from the band edge corresponding to a gap soliton. The
wavefunctions of these bound states are depicted in Figs.
\ref{fig6}b,c. In analogy with the attractive case, we have that
the unstable soliton exists only for a critical value $N_{cr}$ of
the norm and has a behavior similar to the gap-Townes soliton
described before. The decaying property of the repulsive
gap-Townes soliton is clearly illustrated in Fig.\ref{fig7} where
we show the time evolution, as obtained from direct numerical
integrations of Eq.\ref{gpe} (with $g=0, \alpha = 0$), of the
gap-Townes soliton in Fig.\ref{fig6}b (central panel). The left
and right panels of this figure show the decays into the extended
(Bloch) and localized state (gap soliton) when the norm of the
initial condition is, respectively,  slightly decreased or
increased with respect to $N_{cr}$.
Thus, also in this case a gap-Townes soliton appears as a
separatrix between localized and extended states. We remark,
however, that in this case the existence of gap-solitons is a
direct consequence of the band structure (it could not exist
without the OL). The fact that the energy of the Townes soliton is
slightly above the upper band edge is consistent with the
existence of a threshold in the number of atoms (given by the
critical norm of the gap-Townes soliton) below which localized
state of the quintic GPE with OL cannot exist. In the left panel
of Fig. \ref{fig8} we show the delocalizing curve in the plane
$(N, \varepsilon)$ separating localized from extended states. This
behavior is very similar to what reported for repulsive 2D
solitons of the GPE in optical lattices \cite{BS}. In the right
panel of the same figure we show the energies  of the Bloch states
at the top of the band, of the Townes soliton immediately above
the band and of the gap soliton, as a function of $\varepsilon$.
In contrast with the attractive case we see that all energies
monotonically increase with decreasing $\varepsilon$. The absence
of a maximum in the energy curve of the gap soliton at small
values of $\varepsilon$ also indicates the absence of collapse in
this case (compare the right panels of Figs \ref{fig4} and
\ref{fig8}).

\section{Gap-Townes solitons in presence of dissipation}
In the previous sections we have investigated the general
properties of localized states of 1D BEC in OL with  the elastic
three-body interactions modeled by a real quintic nonlinearity. It
is known, however, that  the three-body interactions bears also an
imaginary component corresponding to inelastic collisions which
can be modeled by a damping term of the form $-i \gamma |\psi|^4
\psi$ in the right hand side of Eq. (\ref{gpe}). Recent
theoretical studies \cite{Bed}-\cite{Koh} estimate for $Rb$ a
negative value of the real part of the quintic nonlinearity of the
order of $10^{-26}-10^{-27} cm^6/s$ . From the experiment in Ref.
\cite{Tolra} one can deduce the imaginary part of the three-body
interaction for $Rb$ to be of the order $10^{-30} cm^6/s$. These
numbers imply a ratio between the imaginary and the real part of
this interaction of the order $10^{-3}-10^{-4}$. Other data gives
values of the damping constant of the order $10^{-28} cm^6/s$. Due
to the uncertainty intrinsic in some of these numbers (presently
there are no measurements of the real part of the three-body
interactions) in the following we take the damping constant
$\gamma$ to be a free parameter and investigate both the
underdamped and the overdamped regimes.
\begin{figure}\centerline{
\includegraphics[width=4.cm,height=4.5cm,angle=0,clip]{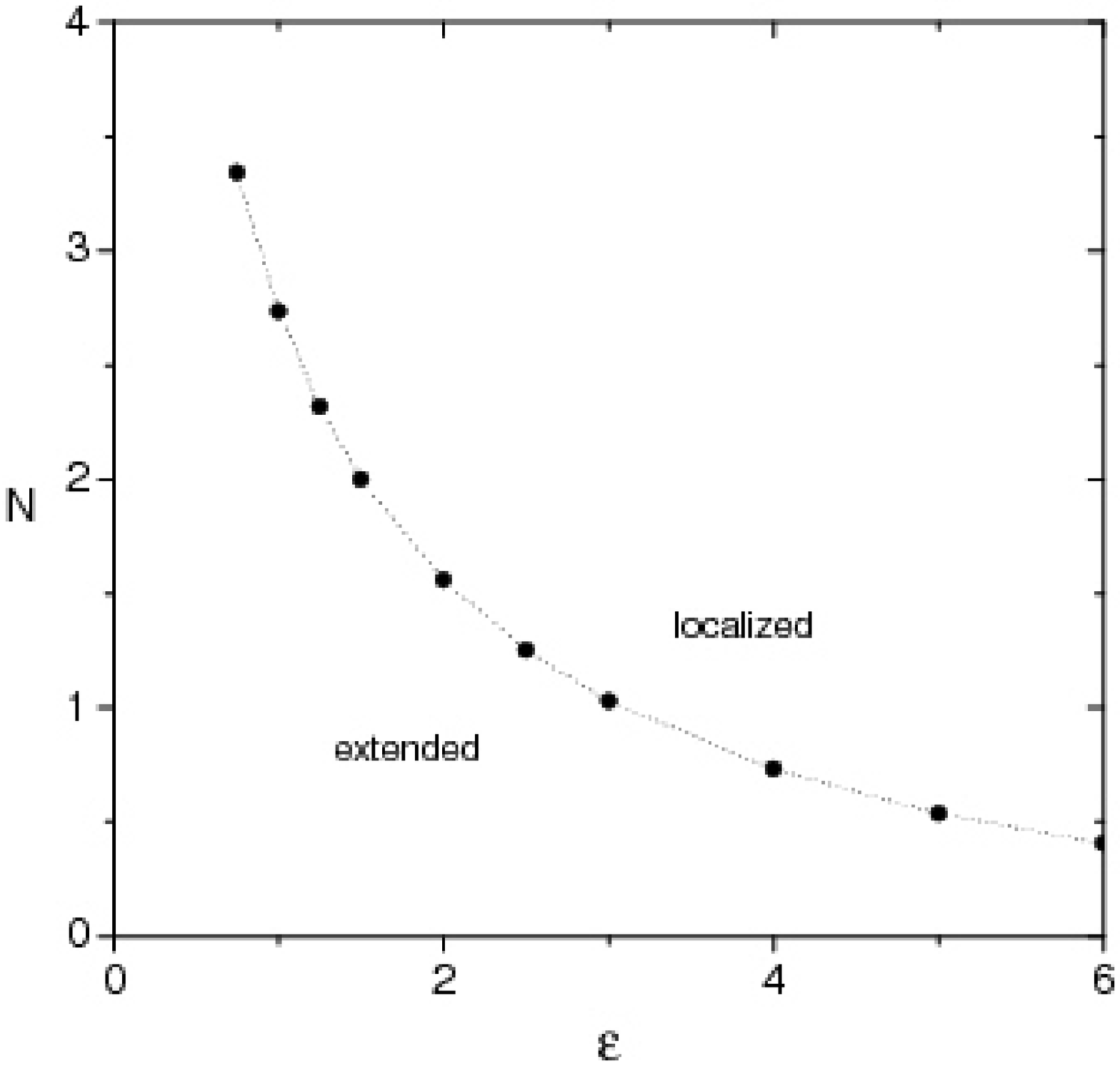}
\includegraphics[width=4.cm,height=4.5cm,angle=0,clip]{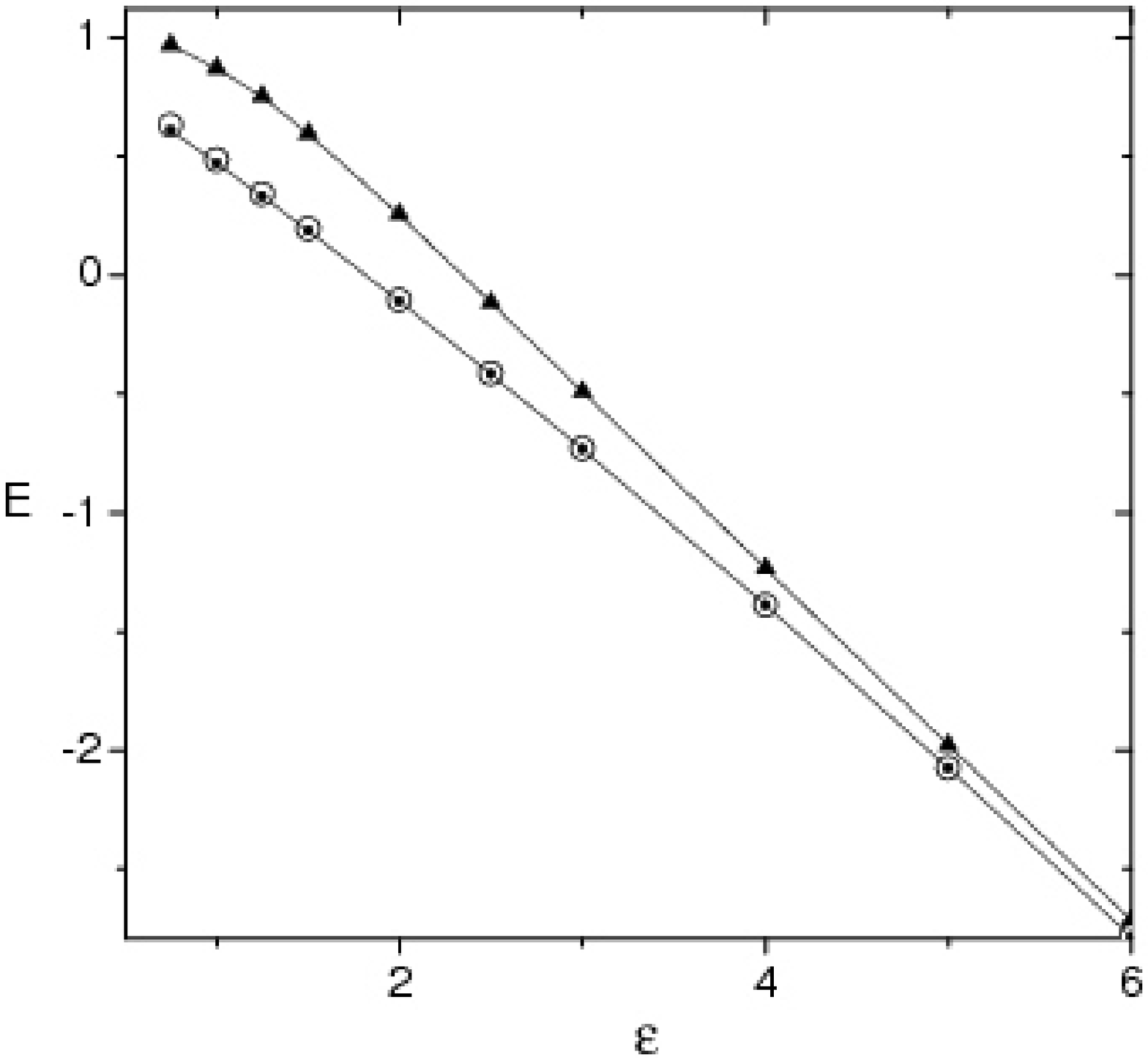}
} \caption{ Left Panel. Delocalizing transition threshold obtained
by direct numerical integrations (filled dots) of the quintic NLS
with $\chi=1.0$. Right Panel. Energies versus $\varepsilon$ of the
gap soliton (top curve with filled triangles), Townes soliton
(curve with filled circles) and Bloch state at the top of the band
(curve with open circles) for the quintic NLS with $\chi=1.0$.
Plotted quantities are in normalized units. } \label{fig8}
\end{figure}

\begin{figure}\centerline{
\includegraphics[width=3.cm,height=5.cm,angle=0,clip]{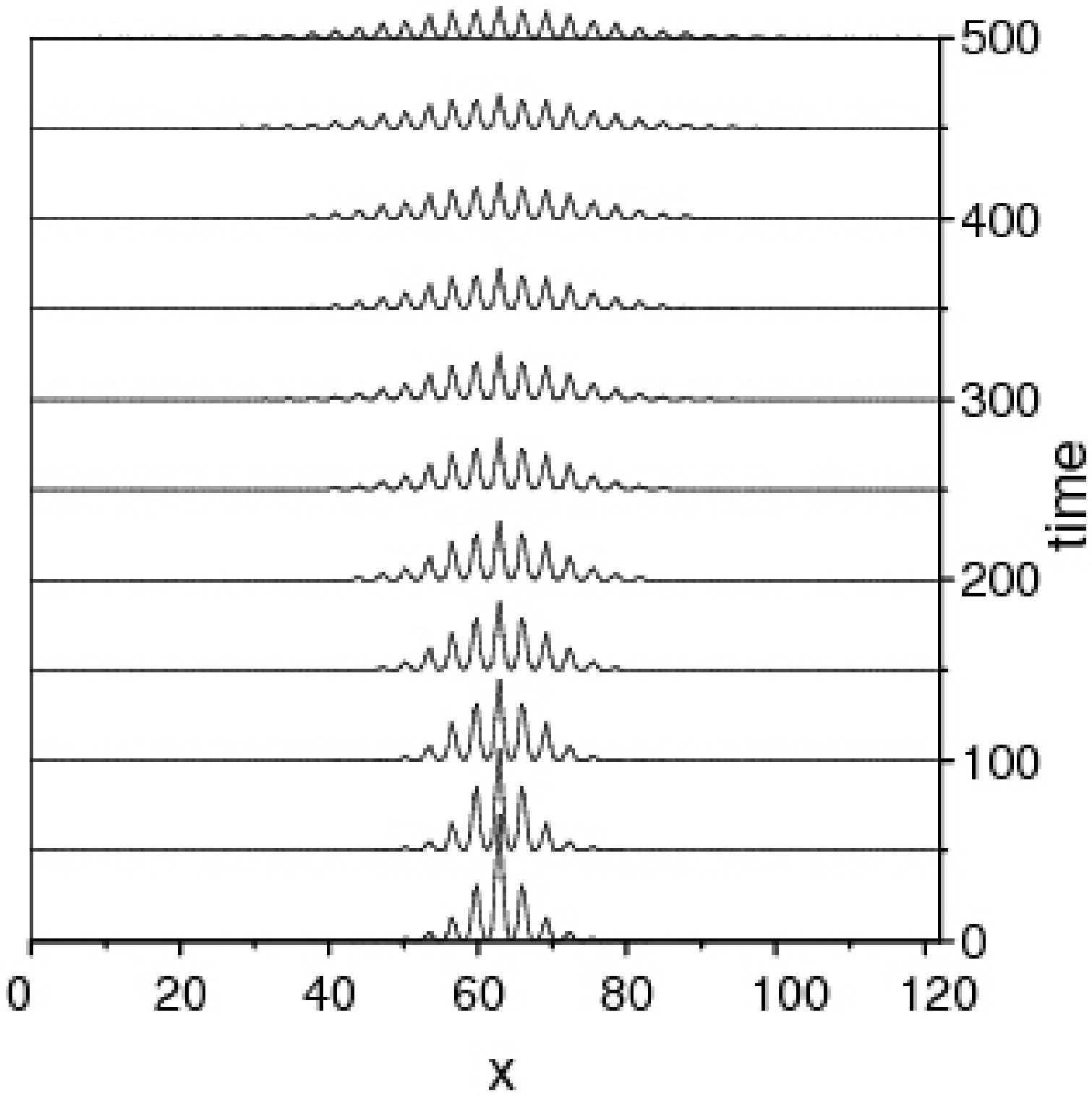}
\includegraphics[width=3.cm,height=5.2cm,angle=0,clip]{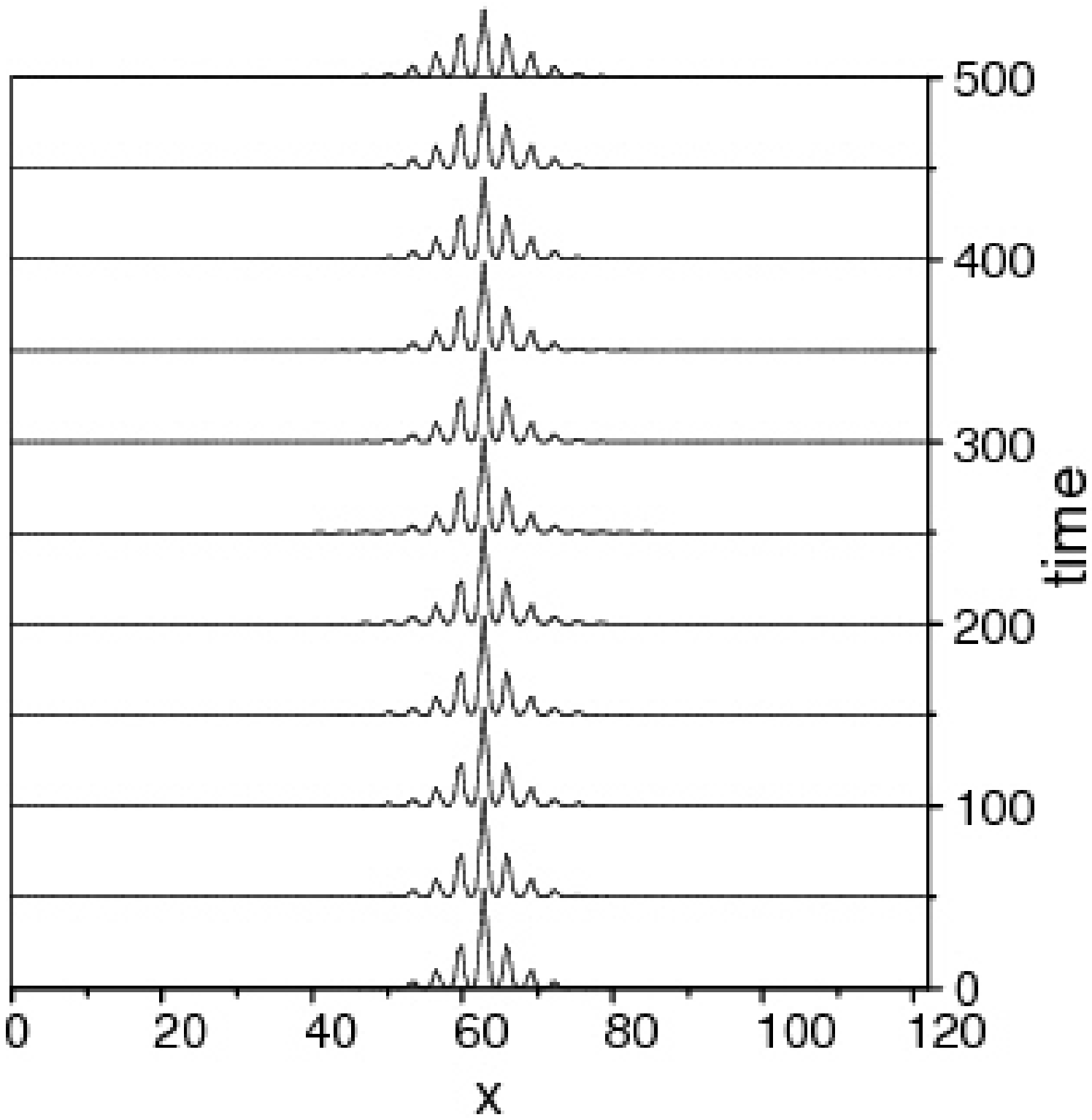}
\includegraphics[width=3.cm,height=5.5cm,angle=0,clip]{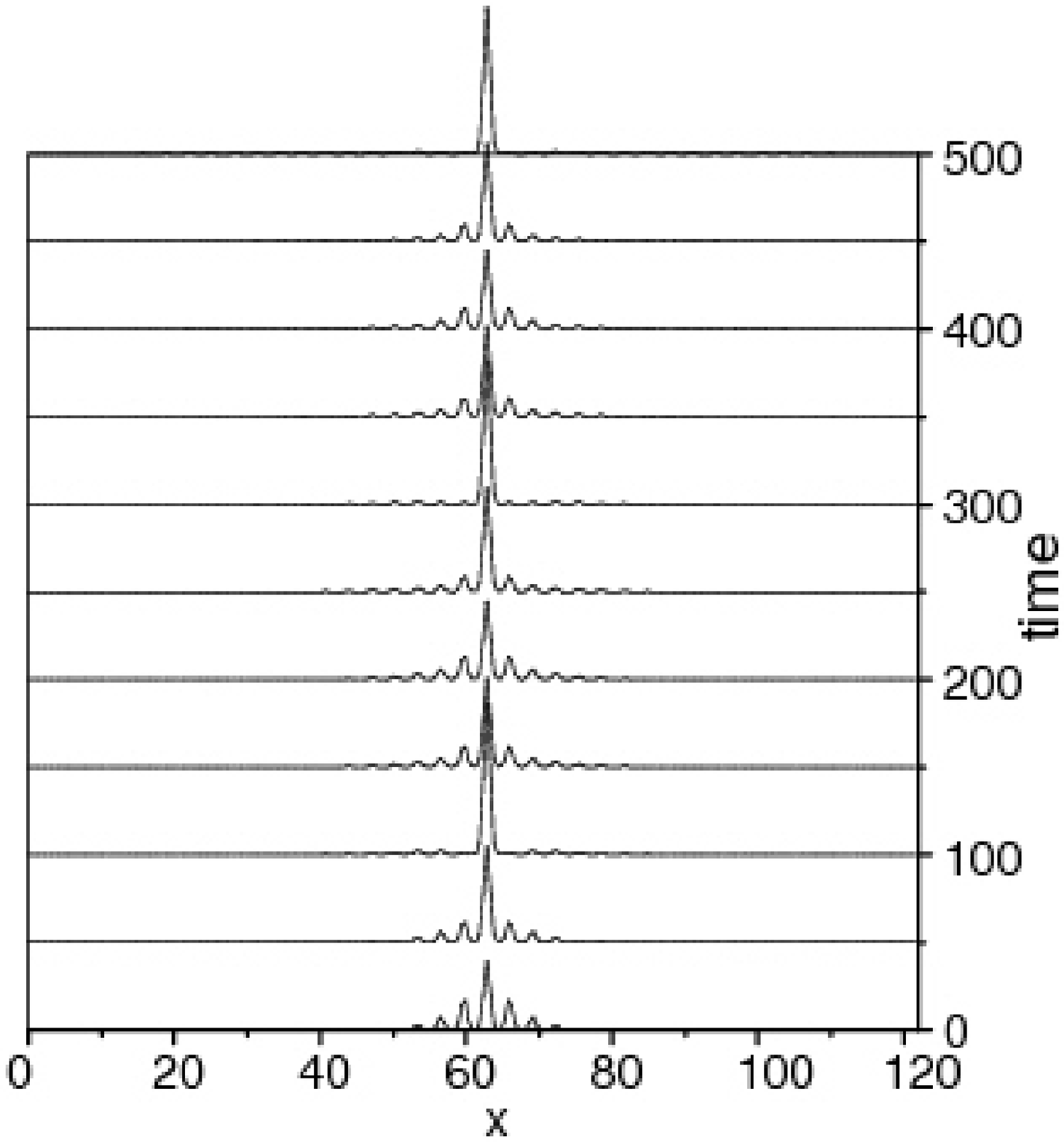}
} \centerline{
\includegraphics[width=3.cm,height=5.cm,angle=0,clip]{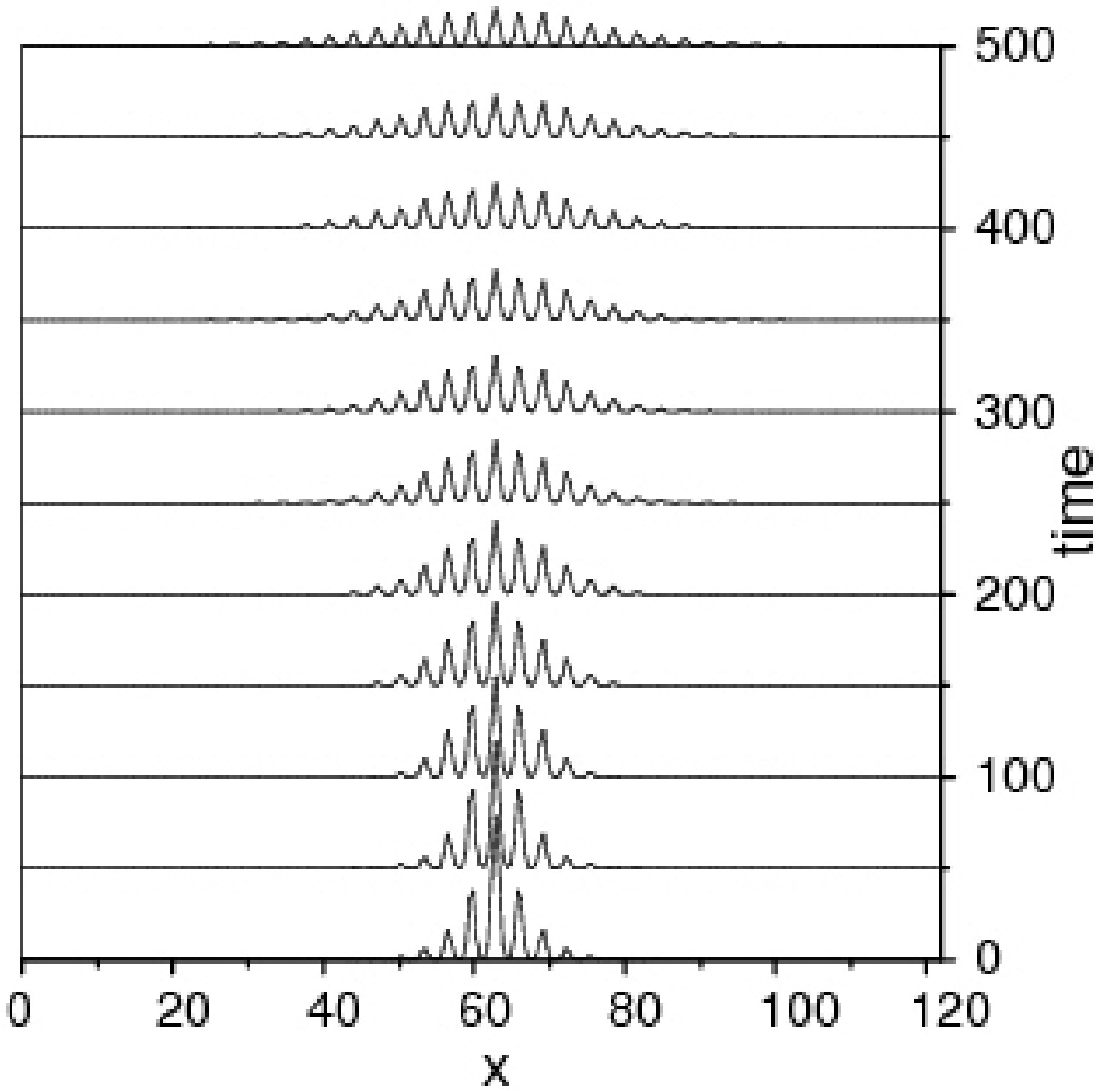}
\includegraphics[width=3.cm,height=5.cm,angle=0,clip]{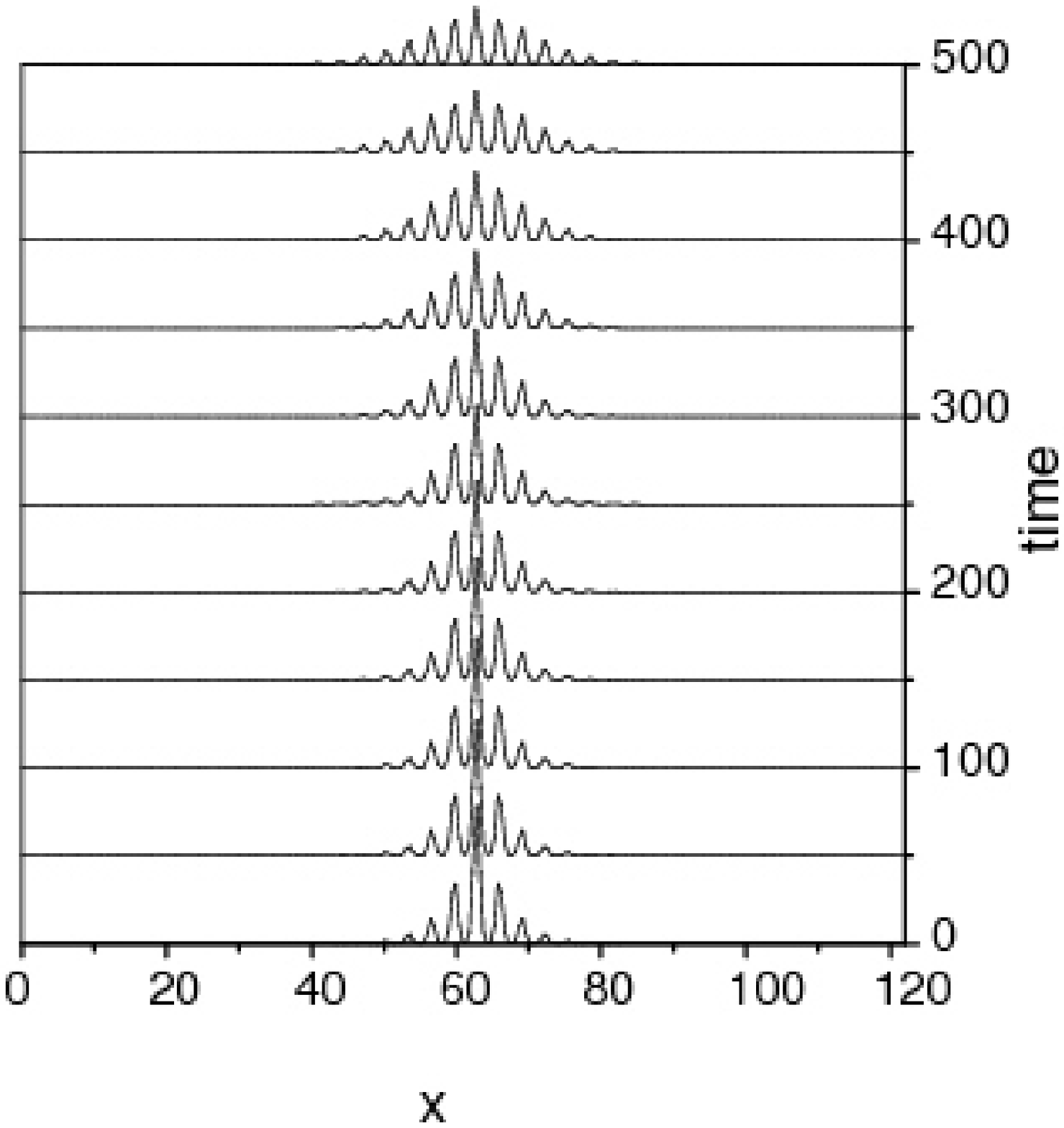}
\includegraphics[width=3.cm,height=5.cm,angle=0,clip]{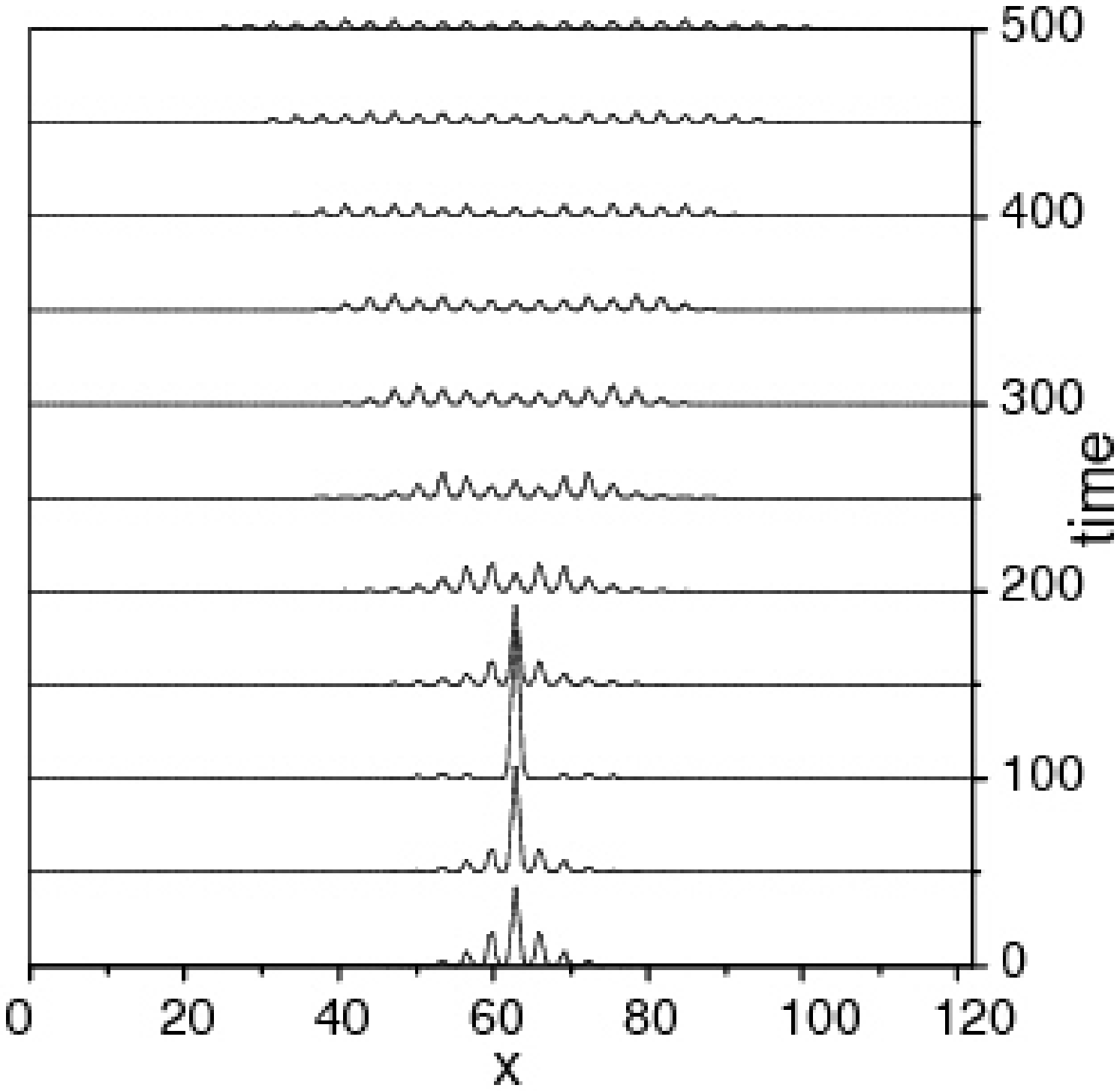}
} \caption{Time evolution of Townes soliton in Fig.\ref{fig2} as
obtained from numerical integration of the dissipative quintic NLS
with attractive interaction. The top three panels refer to three
slightly different values of number of atoms: $N_{cr}=0.87248$
(central panel), $N=1.1 N_{cr}$ (right panel) and $N=0.9 N_{cr}$
(left panel) with a dissipation parameter $\gamma=0.00025$. The
bottom three panels are the same but for $\gamma=0.01$.  Other
parameters are fixed as $\chi=-1$, $\varepsilon = 5$. Plotted
quantities are in normalized units.} \label{fig9}
\end{figure}

The instability of gap-Townes solitons against small fluctuations
in the number of the atoms rises the question of how the scenario
of the previous sections  changes in presence of dissipation. To
answer this question we concentrate on the case of attractive
quintic interactions (qualitatively similar results holds also for
repulsive interactions) in presence of a quintic dissipation. The
results are displayed in Fig. \ref{fig9}. In the top three panels
of this figure we depict the time evolution of the Townes soliton
in Fig. \ref{fig2} in presence of a small damping. We see that in
spite of a slow broadening of the Townes soliton (middle panel),
the situation remains  qualitatively similar to the undamped case
depicted in Fig. \ref{fig3}. Notice that for an overcritical
number of atoms (right top panel) several focusing-defocusing
cycles between the gap-Townes soliton and the gap soliton occur
(these oscillations are induced by the slightly overcritical
starting value of $N$). From  the left panel of this figure we
also see that for an undercritical number of atoms  the decay into
the Bloch states occurs faster than for the zero damping case. At
very long times the gap-Townes soliton in the middle panel of Fig.
\ref{fig9} eventually  decays into the Bloch stated at the edge of
the band.

For larger damping the situation is different, as one can see from
the bottom three panels of Fig.\ref{fig9}. Notice that for an
overcritical value of $N$ the gap-Townes soliton (right panel)
decays into an extended state after performing only one
focusing-defocusing cycle. In this case the dissipation completely
suppresses collapse and localized states  becomes unstable against
decay into the lowest energy Bloch state. It is remarkable,
however, that a signature of the existence of gap-Townes soliton
remains even for moderately large dampings in the focusing
defocusing cycles observed when the decreasing (due to
dissipation) number of atoms crosses the critical value for
existence of  gap-Townes soliton. This fact could indeed be used
in experiments to detect the existence of gap-Townes solitons in
presence of damping.

An analytical explanation of the existence of focusing-defocusing
cycles can be obtained by means of a modified variational analysis
\cite{Bendenson,FAGT} in which the damping is treated as a small
perturbation. In this approach we assume a time dependent
localized state  of the form
\begin{equation}\label{ansatz}
\psi(x,t)= A \exp(-\frac{x^2}{2w^2} + i b x^2 + i\phi ),
\end{equation}
with  parameters  $\eta_i \equiv \{A, w, b, \phi\}$ also time
dependent. The equations of motion for these variables are
obtained from the averaged unperturbed Lagrangian of
Eq.(\ref{gpe}) with $g=0$
\begin{equation}
L=-\frac{\sqrt{\pi}}2A^2 w[w^2 b_t+ 2 \phi_t + \frac 1{w^2}+ 4 w^2
b^2 -2 \varepsilon e^{- w^2} + \frac{2\chi}{3 \sqrt{3}}
A^4]\nonumber
\end{equation}
as
\begin{equation}
\frac{d}{dt}\frac{\partial L}{\partial{\eta_{i,t}}}-\frac{\partial
L}{\partial\eta_i} = \int_{-\infty}^{\infty}dx (R
\psi^{\ast}_{\eta_i} + R^{\ast} \psi_{\eta_i})
\end{equation}
where $R = -i\gamma|\psi|^4 \psi$, and the subscript ${\eta_i}$
denotes the derivative with respect to soliton parameters. From
the above generalized Euler-Lagrange equation one derives the
following equations for the number of atoms $N$ and soliton width
$w$
\begin{eqnarray}\label{dva}
N_{t} &=& -3 \gamma \beta \frac {N^3} {w^2},\nonumber \\
w_{tt} &=& \frac 4{w^3} (1 + \chi \beta N^2) -8\epsilon w
e^{-w^2},
\end{eqnarray}
with $\beta =2/(3\sqrt{3}\pi).$

\begin{figure}\centerline{
\includegraphics[width=3.8cm,height=3.8cm,angle=0,clip]{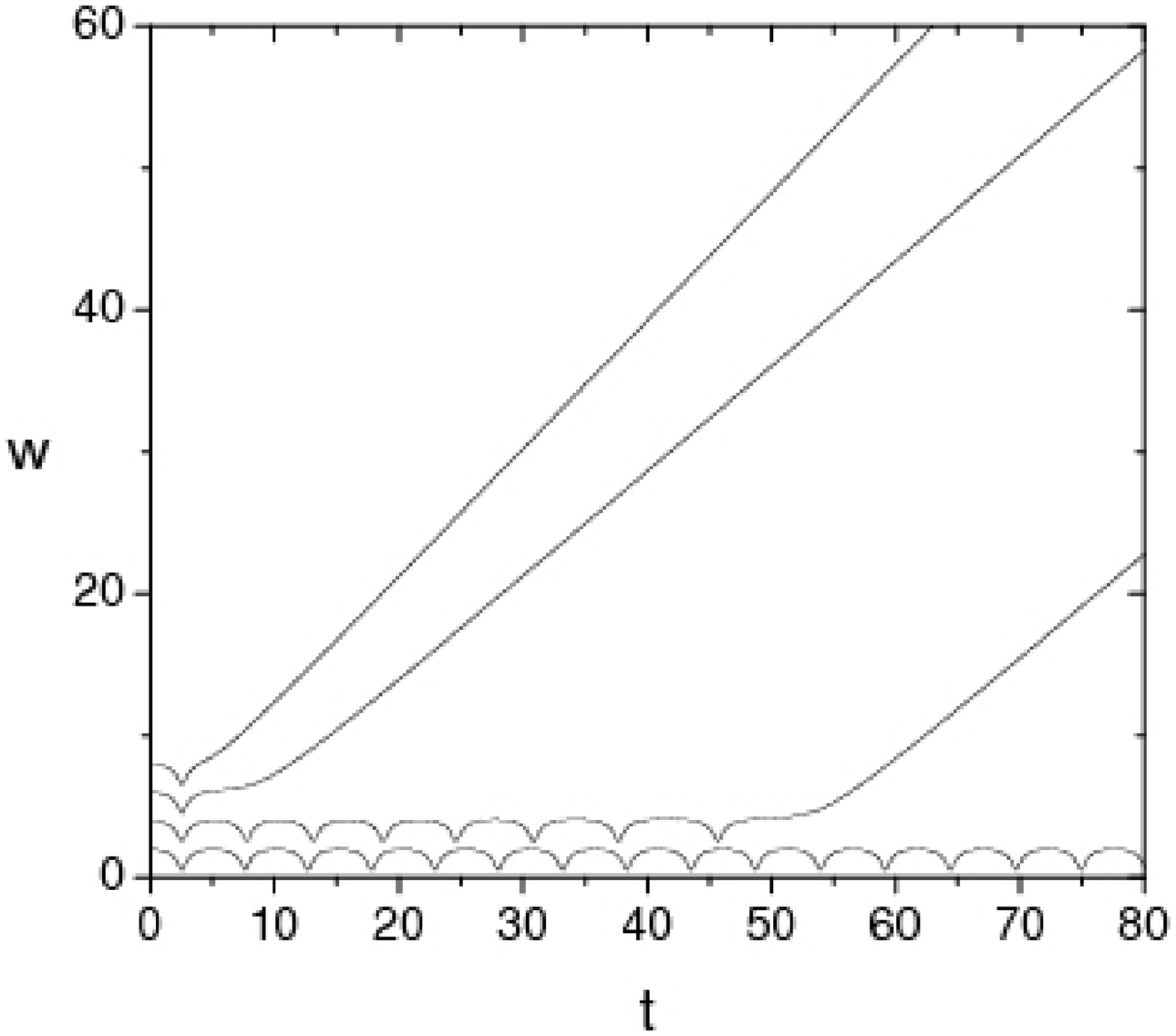}
\includegraphics[width=3.8cm,height=3.8cm,angle=0,clip]{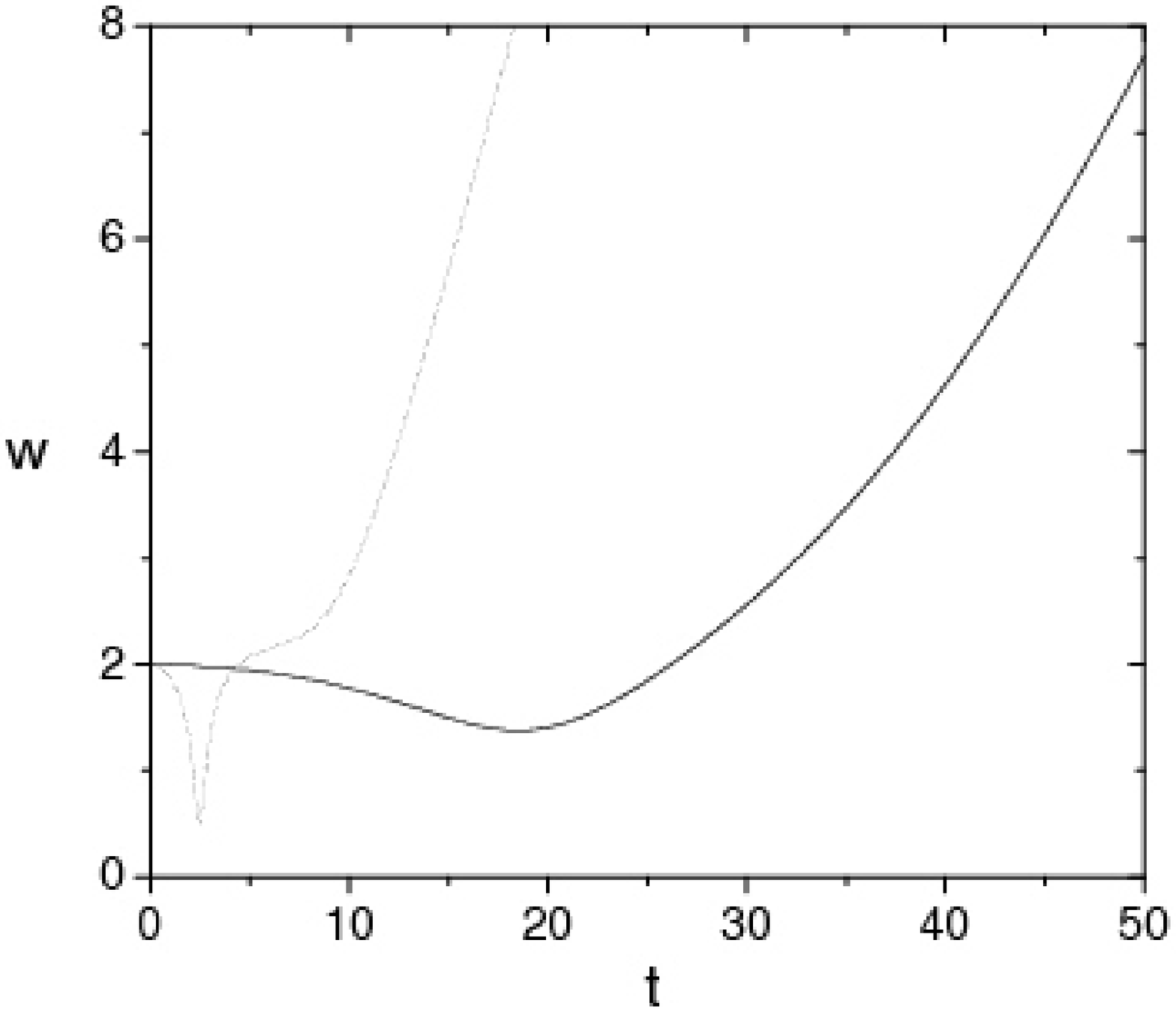}}
\caption{Left panel. Time evolution of the width of the Townes
soliton as obtained from Eqs. \ref{dva} with $\varepsilon=2$,
$\chi=-1$ and damping parameters increased by a factor $10$ going
from the bottom curve ($\gamma=0.0001$) to the top one
($\gamma=0.1$). The last three curves have been shifted
respectively  by $2,4,6$, to avoid overlapping. The initial value
of atoms was detuned from the critical value $N=1.254$
corresponding to the unperturbed Townes by $0.05$. Right panel.
Time evolution of the soliton width obtained from the damped
quintic GPE  (continuous line) and  from the variational equations
\ref{dva} (dotted line). Parameter values are taken as
$\varepsilon=2$, $\chi=-1$, $\gamma=0.01$. The comparison with the
variational results was done by rescaling the soliton width
obtained from the GPE by a factor $11.5$ to have the same initial
condition. Plotted quantities are in normalized units.}
\label{fig10}
\end{figure}

In Fig \ref{fig10} we show the time evolution of the soliton width
for different values of the dissipation constant. We wee that for
very low dampings the width oscillates around a constant value,
meaning  that the localized state undergoes periodic
focusing-defocusing cycles in agreement with the right top panel
of  Fig. \ref{fig9}. By increasing the damping we see that after a
certain number of cycles,  the soliton width starts to grow
monotonically signaling the decay of the gap soliton  into an
extended state. Notice that even for relatively high dampings
there is one focusing-defocusing cycle, this being in good
qualitative agreement with what observed in the bottom right panel
of Fig. \ref{fig9}. In the right panel of Fig. \ref{fig10} we make
a quantitative comparison between  the dynamics of the soliton
width as obtained from the VA and from the damped quintic NLS.
From this figure we see that although the main feature of the
phenomenon is correctly captured (i.e. existence of only one
focusing-defocusing cycle before decay), the variational equations
are not in good quantitative agreement with the numerical
integration of the full system. This may be due to the fact that
the ansatz in Eq.(\ref{ansatz}) is more appropriate for strongly
localized states than for extended gap-Townes ones. It is
interesting, however, that for $\epsilon = 0$ the variational
equations become similar in form to the ones considered for the 2D
cubic NLSE in presence of a cubic dissipation
$-i\gamma|\psi|^{2}\psi$ \cite{Fibich}. In this case one can show
that the equation for $w$ reduces to a modified Airy equation for
which a focusing-defocusing cycle always exists. When $\epsilon
\neq 0$, Eq.s (\ref{dva}) predict that for decreasing $\gamma$ the
number of focusing-defocusing cycles grows, this being  in good
agreement with direct numerical simulations. Moreover, from the
left panels of Figs. \ref{fig4} and \ref{fig8} we see that the
critical value of the number of atoms for existence of gap-Townes
soliton decreases by increasing $\varepsilon$, this meaning that
in strong optical lattices the effect of the quintic damping on
gap-Townes solitons is effectively reduced (due to the reduced
norm), this being confirmed also by numerical simulations.
\begin{figure}\centerline{
\includegraphics[width=3.8cm,height=4.2cm,angle=0,clip]{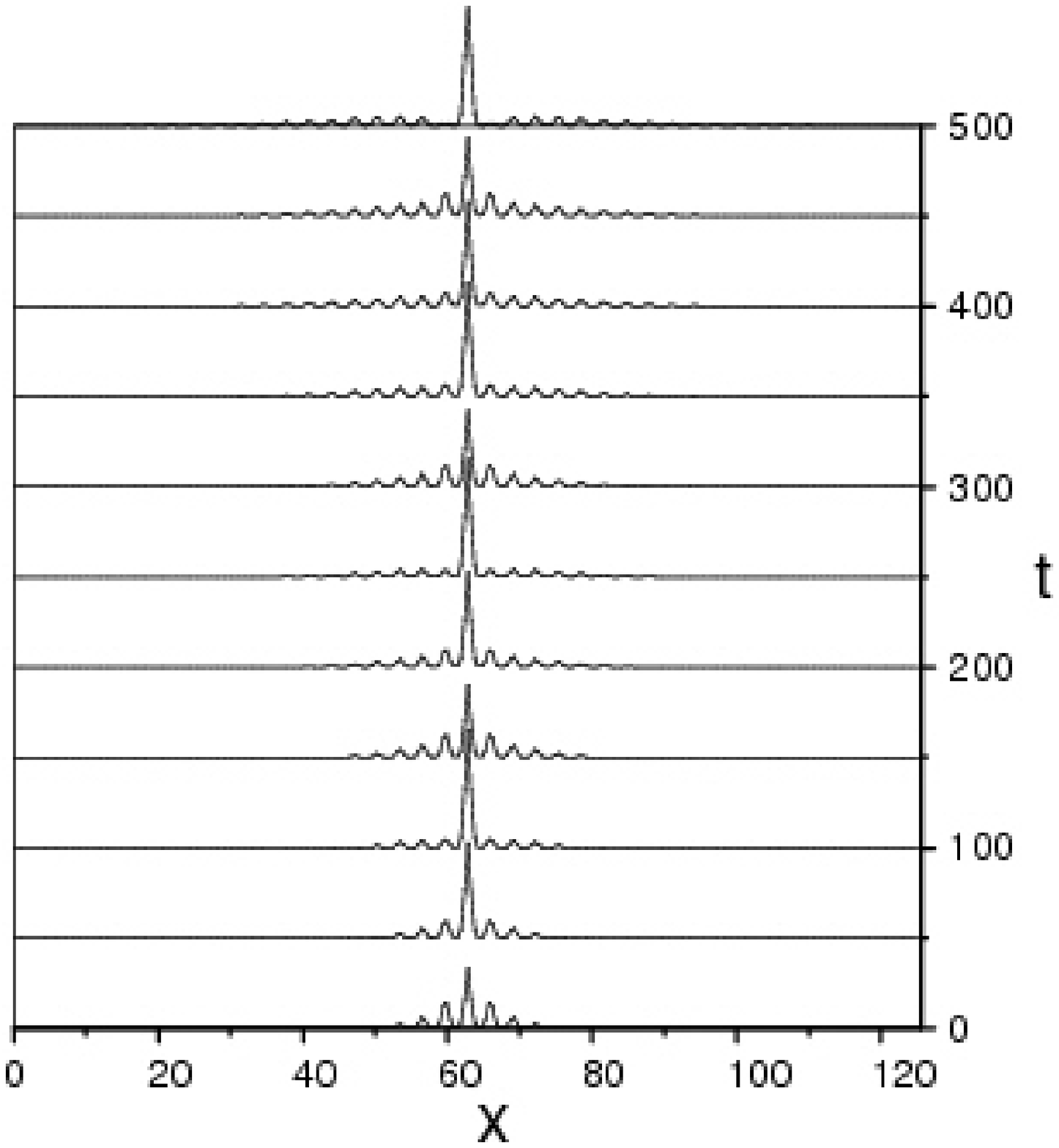}
\includegraphics[width=3.8cm,height=3.8cm,angle=0,clip]{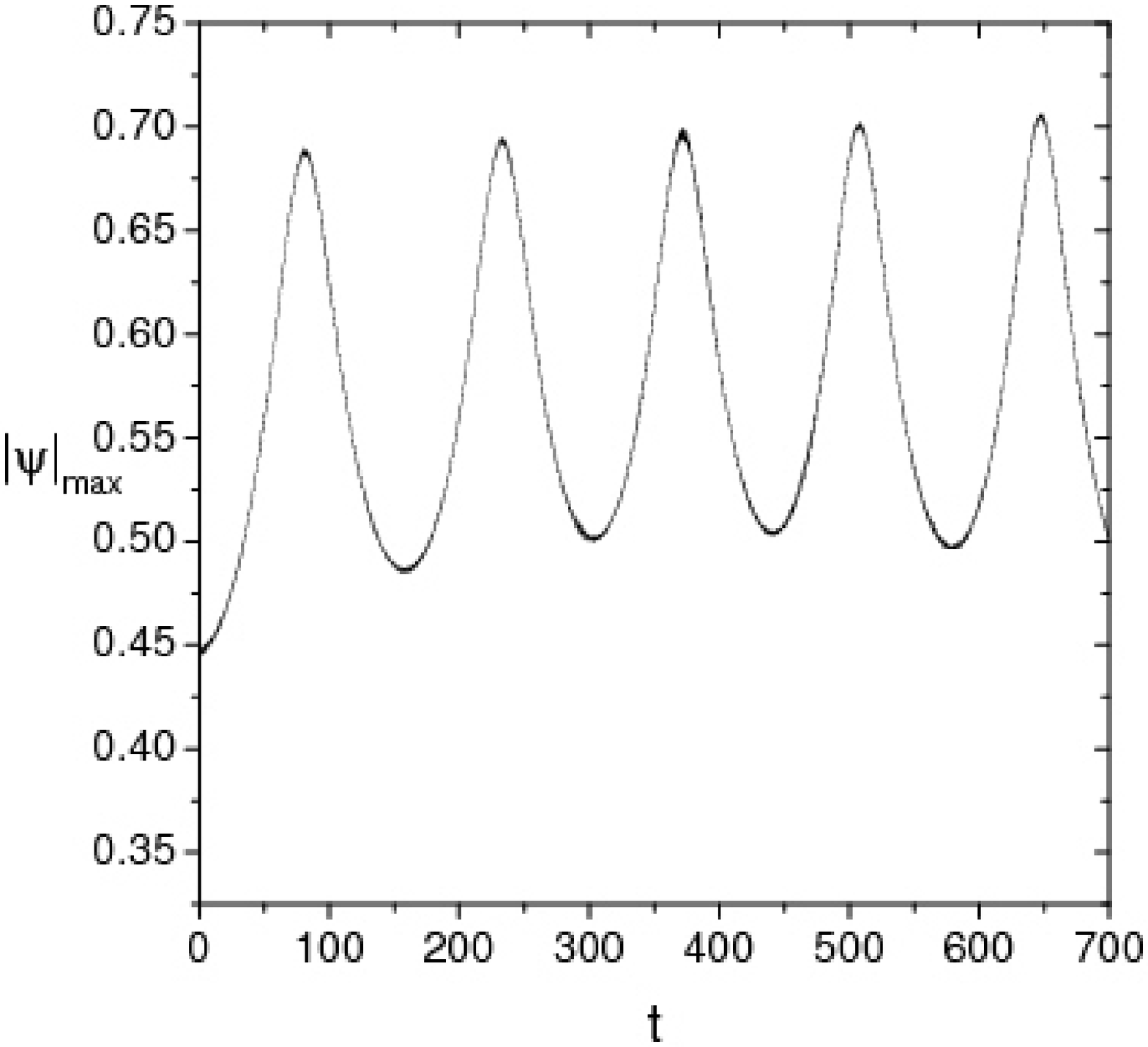}}
\caption{Left panel. Time evolution of the gap-Townes soliton in
the lower right panel of Fig. \ref{fig9} in presence of  quintic
dissipation $\gamma=0.01$ and linear amplification of strength
$\delta=8.95 \cdot 10^{-4}$. Right panel. Time evolution of the
gap-Townes soliton amplitude in the left panel. Plotted quantities
are in normalized units.} \label{fig11}
\end{figure}

We remark that besides the decay phenomena induced by the quintic
dissipation one could also consider the feeding of atoms in the
condensate from the thermal cloud \cite{kagan}.  The injection of
matter in the condensate can be modeled with a linear imaginary
term (of sign opposite to $\gamma$) in the NLS equation of the
form $i \delta \psi $. This term can balance the damping and
stabilize localized states against decay even in the overdamped
regime, this leading to the formation of dissipative gap solitons.
We illustrate this phenomenon in Fig. \ref{fig11} where the time
evolution of the slightly overcritical gap-Town soliton shown in
the bottom right panel of Fig. \ref{fig9} is reported for a linear
amplification term of strength $\delta=8.95 \cdot 10^{-4}$. From
this figure we see that the localized state instead of decaying
into the Bloch state at the bottom of the band (see Fig.
\ref{fig9}), it executes breather-like oscillations (i.e.
focusing-defocusing cycles) with emission of radiation. In terms
of band picture this breathing state can be thought as a non
stationary states with energy oscillating between the gap-Townes
soliton and the gap soliton in Fig. \ref{fig2}a. We believe this
being a general mechanism for existence of breather-like
excitations in 1D BEC in OL in presence of damping and
amplification.

\section{Conclusions}
In this paper we have used the  periodic 1D NLS equation with
quintic nonlinearity to investigate localized states in 1D BEC in
OL with three-body interactions. The existence of unstable
solitons similar to the usual Townes soliton of the cubic 2D NLS
has been established. These states have been shown to have energy
located in the forbidden zones of the band structure, very close
to band edges, separating decaying states from stable localized
ones (gap-solitons). The existence of gap solitons appears to be a
mechanism for arresting collapse in attractive low dimensional BEC
with three-body interactions in OL and the analysis of gap-Townes
solitons appears to be important for characterizing the
delocalizing transition in these systems. To this regard we remark
that the region of existence of localized states in the
$N,\varepsilon$ plane is bounded by two limiting curves. The first
curve separates extended states from localized ones and therefore
characterizes the delocalizing transition. This curve coincides
with the curve of existence of gap-Towenes solitons. For the 1D
quintic NLS with periodic potential the delocalizing transition
exists  for both attractive and repulsive interactions.
Delocalizing curves have been investigated in Fig.s \ref{fig4},
\ref{fig8}. A similar behavior is expected to be valid also for
the standard GPE with periodic potential in higher dimensions. The
second curve separates the localized solutions from the collapsing
ones, giving an upper bound for the existence of gap-solitons.
This curve obviously exists only for attractive interactions (for
repulsive interactions a negative effective mass, although
producing  an attraction among atoms which leads to the formation
of gap solitons, is usually insufficient to induce collapse). Due
to the numerical difficulties involved in the study of collapsing
solution, the threshold curves for collapse have not been
investigated in this paper. For this it will be useful to develop
an analytical approach to provide estimates of the critical values
for collapse \cite{Kon2}.

We have also investigated the influence of a dissipative imaginary
part of the three-body interaction on gap-Townes solitons. In
particular, we have shown that the damping introduces effects
which are important when the ratio between the imaginary and the
real part of the three-body interaction is not small. For the case
of rubidium this ratio is expected to be small and the situation
described for the underdamped case should be qualitatively valid.
It is remarkable, however, that even for moderate large damping a
signature of the existence of gap-Townes solitons remains in the
focusing-defocusing cycles observed when the number of atoms
crosses the critical value for their existence.

Finally, we have shown that the presence of a small linear
amplification term into the damped quintic NLS equation, modelling
the feeding of atoms from the thermal cloud, can stabilize gap
solitons into breather-like excitations even in presence of large
damping.

The results of this paper suggest the possibility to
experimentally observe gap-Townes solitons in low dimensional BEC.

\acknowledgments \noindent The authors acknowledge L.Pitaevskii,
V.V. Konotop, B.B. Baizakov and V. Korepin for interesting
discussions. FKhA wishes to thank the Department of Physics
"E.R.Caianiello" and the University of Salerno  for a ten months
research grant during which this work was done. MS acknowledges
partial financial support from a MURST-PRIN-2003 Initiative.

\end{document}